\pdfoutput=1
\documentclass[12pt]{article}
\usepackage[top=1in,bottom=1in,left=1in,right=1in]{geometry}
\usepackage{graphicx}
\usepackage{amsfonts, amssymb, amsmath, amsthm, booktabs, dsfont, float, xr-hyper, hyperref, mathrsfs, euscript, enumitem, bm, bbm, pifont, multirow, tabularx, multicol, tikz}
\usepackage[onehalfspacing]{setspace}
\usepackage[labelfont=bf,textfont=it]{caption} 
\usepackage{subcaption}
\hypersetup{pdfborder = {0 0 0},colorlinks=true,allcolors=blue}
\usepackage{natbib}
\usepackage[title]{appendix}
\usepackage{placeins}
\usepackage{threeparttable}
\usepackage[normalem]{ulem}
\usepackage{algorithm}
\usepackage{algpseudocode}

\newtheorem{thm}{Theorem}

\newtheorem{assumption}{Assumption}

\numberwithin{equation}{section}

\theoremstyle{definition}
\newtheorem{remark_tmp}{Remark}

\theoremstyle{definition}
\allowdisplaybreaks[1]

\renewcommand*{\arraystretch}{.6}

\renewcommand{\P}{\mathbb{P}}
\newcommand{\E}{\mathbb{E}}
\newcommand{\V}{\mathbb{V}}

\newcommand{\indic}{\mathbbm{1}}

\newcommand{\bA}{\mathbf{A}}

\newcommand{\bW}{\mathbf{W}}

\newcommand{\bZ}{\mathbf{Z}}

\newcommand{\bw}{\mathbf{w}}

\newcommand{\bgamma}{\boldsymbol{\gamma}}

\newcommand{\bxi}{\boldsymbol{\xi}}

\newcommand{\blambda}{\boldsymbol{\lambda}}

\newcommand{\bOO}{\boldsymbol{0}}

\newcommand{\ahat}{\widehat{\alpha}}

\newcommand{\thetahat}{\widehat{\theta}}

\newcommand{\select}{\boldsymbol{\mathsf{s}}}

\DeclareMathOperator{\R}{\mathbb{R}} 					 

\usepackage{listings}
\begin{document}

\title{\vspace{-.25in}\textbf{Treatment Effect Heterogeneity\\ in Regression Discontinuity Designs}\thanks{We thank participants at various seminars and conferences for comments. Cattaneo and Titiunik gratefully acknowledge financial support from the National Science Foundation (SES-2019432 and SES-2241575).}\bigskip}
    \author{Sebastian Calonico\thanks{Graduate School of Management, UC Davis.} \and
        Matias D. Cattaneo\thanks{Department of Operations Research and Financial Engineering, Princeton University.} \and
	    Max H. Farrell\thanks{Department of Economics, UC Santa Barbara.} \and
	    Filippo Palomba\thanks{Department of Economics, Princeton University.} \and
	    Roc{\'i}o Titiunik\thanks{Department of Politics, Princeton University.}}
\maketitle

\begin{abstract}
    Empirical studies using Regression Discontinuity (RD) designs often explore heterogeneous treatment effects based on pretreatment covariates, even though no formal statistical methods exist for such analyses. This has led to the widespread use of ad hoc approaches in applications. Motivated by common empirical practice, we develop a unified, theoretically grounded framework for RD heterogeneity analysis. We show that a fully interacted local linear (in functional parameters) model effectively captures heterogeneity while still being tractable and interpretable in applications. The model structure holds without loss of generality for discrete covariates. Although our proposed model is potentially restrictive for continuous covariates, it naturally aligns with standard empirical practice and offers a causal interpretation for RD applications. We establish principled bandwidth selection and robust bias-corrected inference methods to analyze heterogeneous treatment effects and test group differences. We provide companion software to facilitate implementation of our results. An empirical application illustrates the practical relevance of our methods.
\end{abstract}


\thispagestyle{empty}
\clearpage

\doublespacing
\pagestyle{plain}


\section{Introduction}
    \label{sec:intro}

Analyzing heterogeneous treatment effects based on pretreatment covariates is crucial in modern causal inference. While the average causal effect provides a broad measure of a treatment's effectiveness and informs general policy decisions, uncovering patterns of covariate heterogeneity is often essential for evaluating substantive hypotheses and specific policy interventions. Covariate-based heterogeneous treatment effects can help understand fairness and differential impacts across subpopulations, develop targeted interventions, and optimize social policies.

Over the past decade, the regression discontinuity (RD) design has become a prominent tool for causal inference, leading to extensive methodological advancements for identification, estimation, and inference for average treatment effects in various contexts. Despite its widespread adoption, there is still a notable lack of rigorous methods for studying heterogeneity based on pretreatment covariates, resulting in many ad-hoc empirical approaches. We address this gap by developing practical, theoretically-grounded tools for heterogeneity analysis in RD designs, putting forward a unified methodology for empirical researchers studying heterogeneous treatment effects based on pretreatment covariates.

Extending RD estimation and inference to encompass heterogeneous treatment effects presents a high-dimensional, nonparametric challenge: while theoretically feasible, it is often impractical without model restrictions. Consequently, empirical researchers typically use semiparametric models and explore heterogeneity in limited dimensions. Following empirical practice, we study the most common approach for RD treatment effect heterogeneity analysis: local least squares regression with linear-in-parameters interactions with pretreatment covariates. This model balances practical applicability with the flexible, nonparametric nature of the RD design.

Our first contribution is to clarify the conditions under which a local regression with linear interactions accurately recovers meaningful causal heterogeneous RD treatment effects. We show that heterogeneous effects are identifiable when potential outcomes follow a local linear-in-parameters functional coefficient model in the pretreatment variables. This structure is both flexible and interpretable, and is without loss of generality for binary (and orthogonal) covariates. Without this structure, the linear interactions model leads to a less interpretable and less useful probability limit, as the nonparametric nature of the estimation method hampers its usual causal interpretation emerging from the best linear prediction viewpoint. Therefore, from a practical perspective, we show that (i) heterogeneity analysis based on local polynomial regression methods with interactions is fully warranted without additional assumptions when the pretreatment covariates are binary orthogonal variables (e.g., indicator variables for each mutually exclusive level of a categorical variable), but (ii) heterogeneity analysis requires a semiparametric identifying assumption in the general case of discrete or continuous covariates for causal inference.

We then present formal methods for estimation and inference of heterogeneous RD treatment effects, developing optimal bandwidth selection, point estimation, and robust bias-corrected inference. Our key large-sample econometric results include consistency, a mean squared error expansion, a central limit theorem, and standard error estimators robust to both heteroskedasticity and clustering. We show that the optimal bandwidth depends on the specific inference target, implying different optimal bandwidths for average effects, subgroups, and group difference tests. Our goal is to standardize empirical practice, where bandwidth selection is often ad-hoc, particularly for heterogeneity analysis. From that perspective, we identify conditions under which a single bandwidth can be validly used for the entire analysis, thereby enhancing the practical appeal of our proposed methods.

Despite its major role in empirical work, covariate adjustment for heterogeneity analysis in RD designs has been minimally explored. The only antecedents we are aware of are \cite{hsu2019testing,Hsu-Shen_2021_JAE}, who use moment condition methods to test for the presence of covariate-based heterogeneity, and \cite{Reguly2021-wp} and \cite{alcantara2025LearningConditionalAverage}, who leverage machine learning methods to recover nonparametric covariate-based heterogeneity. None of these contributions study the properties of the most common approach in empirical work: semi-linear local polynomial regression with pretreatment covariate interactions. We address this gap by setting up a standard, tractable local least squares framework with covariate interactions, and then providing novel identification, estimation, and robust bias-corrected inference methods to support heterogeneity analysis in RD applications.

By introducing a comprehensive framework for identification, estimation, and inference in RD designs, we aim to unify the study of heterogeneous treatment effects in empirical research. The absence of such a framework has led to a variety of empirical approaches---some valid, others not---creating inconsistencies and confusion in the literature. To illustrate this issue, Section SA3 in the supplemental appendix compiles a sample of papers from AEA journals over the past decade that analyze covariate heterogeneity in RD settings, provided they offer sufficient methodological details. While not exhaustive or necessarily representative, this review reveals three common patterns. First, most studies examine heterogeneity for binary covariates, but often by discretizing categorical or continuous variables. Second, a single bandwidth, often selected in an ad hoc manner, is typically used for both average and heterogeneous treatment effects analysis. Third, and most importantly, inference is rarely conducted properly, regardless of bandwidth choice or covariate granularity. Our proposed methods maintain the simplicity and interpretability of standard empirical practice while introducing objective procedures that bring validity and rigor.

We illustrate our methodological results by revisiting the analysis in \cite{Akhtari-Moreira-Trucco2022_AER}, who study how political turnover in mayoral elections in Brazil affects public service provision. The analysis exploits close elections, a canonical RD setting, using a rich data set of Brazilian mayoral elections. \cite{KlasnjaTitiunik2017-APSR} originally analyzed these data and RD setting, documenting a negative incumbency advantage; see further references therein. We begin by replicating the results in \cite{Akhtari-Moreira-Trucco2022_AER}, who considered both standard RD treatment effects and a limited measure of covariate-heterogeneity based on municipality-level monthly household income. We recover their point estimates of heterogeneity for income dichotomized as above or below the median, and also report valid uncertainty quantification; our results are qualitatively in line with theirs. Then, we use our framework to study income discretized by quartile and by decile, revealing further nuance in the heterogeneity of treatment effects by income level. To conclude, we also show that using income directly (as a continuous variable) provides an accurate summary of the heterogeneity in this application, which is (i) highly compatible with the discretized estimates, (ii) directly interpretable, and (iii) more efficient for estimation and inference. For implementation, we consider two local linear-in-parameters semiparametric heterogeneity models: one taking income linearly, and the other quadratically. This empirical exercise demonstrates how our methods can be used to enhance RD heterogeneity analysis and highlights the importance of clearly understanding how to best capture heterogeneity in the econometric model.

Our paper contributes to the methodological literature on RD designs \citep[see][for a review]{Cattaneo-Titiunik_2022_ARE}, shedding new light on the role of pretreatment covariates for identification, estimation, and inference of heterogeneous RD treatment effects. \citet{Cattaneo-Keele-Titiunik_2023_HandbookCh} overviews covariate adjustment in RD designs and discusses its three most useful roles: (i) increasing estimation efficiency, (ii) changing the parameter of interest, and (iii) heterogeneity analysis. For efficiency gains, \cite{Calonico-Cattaneo-Farrell-Titiunik_2019_RESTAT}, \cite{Ma_Yu2024_WP}, \cite{Arai-Otsu-Seo_2025_ET}, and references therein, give results leveraging pretreatment covariates to improve precision in the estimation of the average RD treatment effect. There is also a rich literature on the inclusion of covariates for identification of other RD treatment effects: for example, see \cite{AngristRokkanen2015-JASA} and \cite{Cattaneo-Keele-Titiunik-VazquezBare_2021_JASA} for extrapolation, \citet{Peng-Ning_2021_PMLR} and \cite{Caetano-Caetano-Escanciano2024_WP} for weighted average treatment effects, \citet{Grembi-Nannicini-Troiano_2016_AEJ-Applied} for difference-in-discontinuities, and \citet{Frolich-Huber_2019_JBES} for nonparametric methods, among many others. As mentioned, the role of covariates for heterogeneity analysis has received the least attention in the literature, and our paper is the first to study it in the context of local semi-linear regression with interactions, the most common approach used in practice. Finally, covariates are sometimes used in an attempt to ``fix'' a ``broken'' RD design. This is not valid without strong assumptions: it is analogous to employ covariate adjustment to ``fix'' a ``broken'' randomized assignment in experimental settings. Our paper does not speak to this practice, and we recommend against it. See \citet{Cattaneo-Keele-Titiunik_2023_HandbookCh} for further discussion.

The rest of the paper proceeds as follows. Section \ref{sec:setup} gives details on the RD setup and estimation. Our main technical results are summarized in Section \ref{sec:theory}, and we present an empirical illustration in Section \ref{sec:application}. Section \ref{sec:conclusion} concludes. The supplemental appendix contains more general, technical results, all proofs, implementation details, and additional numerical results. The software package \texttt{rdhte} \citep{Calonico-Cattaneo-Farrell-Palomba-Titiunik_2025_Stata} provides a general-purpose implementation of our methodological results and is available at \url{https://rdpackages.github.io/rdhte/}.

\section{Setup}
    \label{sec:setup}

We consider the standard sharp regression discontinuity design setting with a continuous running variable \citep[see][for a practical introduction]{Cattaneo-Idrobo-Titiunik2019_book,Cattaneo-Idrobo-Titiunik2023_book}. The defining characteristic is that a binary treatment $T_i \in \{0,1\}$ is assigned to each unit based on whether the continuous running variable $X_i$ is above or below a known cutoff, denoted by $c$. Thus, $T_i = \indic\{X_i \geq c\}$. The observed outcome is $Y_i = T_i Y_i(1) + (1-T_i) Y_i(0)$, where $Y_i(t)$ is the potential outcome for unit $i$ under treatment $T_i = t$. In addition to $Y_i$, $T_i$, and $X_i$, we observe a vector $\bW_i \in \R^d$ of \emph{pretreatment} covariates to be used for heterogeneity analysis. Distinct from $\bW_i$, we may have an additional set of covariates $\bZ_i\in\R^{d_z}$ used for efficiency purposes only; see below. We assume a random sample $(Y_i, T_i, X_i, \bW_i, \bZ_i)$, for $i=1,\dots,n$, is observed.

The usual parameter of interest in sharp RD designs is the average effect of $T_i$ on $Y_i$ at the cutoff. To simplify notation, we set the cutoff to be $c=0$. Then, the average RD treatment effect is
\begin{equation}
    \label{eqn:ate}
    \tau = \E[Y_i(1) - Y_i(0) | X_i = 0].
\end{equation}
Identification of $\tau$ follows from standard continuity assumptions \citep{Hahn-Todd-VanderKlaauw2001_Ecma}, while estimation and inference are typically done with local linear regression and robust bias correction \citep{Calonico-Cattaneo-Titiunik_2014_ECMA}. More precisely, the point estimate $\dot{\tau}$ of the causal parameter $\tau$ is the coefficient on $T_i$ in a least squares regression of $Y_i$ on a constant, $T_i, X_i$, and $T_i \cdot X_i$, using only the observations such that $|X_i - 0| \leq h$, for a bandwidth $h$, and weighted by $K((X_i - 0)/h)$, for a kernel function $K(\cdot)$. This fit, and those below, are defined formally in the Supplemental Appendix (Section SA1.2). We denote the resulting estimated equation by
\begin{equation}
    \label{eqn:reg 1}
    \dot{Y}_i = \dot{\mu}  + \dot{\tau} T_i + \dot{\omega}_1 X_i + \dot{\omega}_2 T_i X_i.
\end{equation}
The coefficient $\dot{\tau}$ is identical to the difference in intercepts from fitting separate (local weighted) least squares regressions on either side of the cutoff. Conceptually, it is important to remember that \eqref{eqn:reg 1} is formally a nonparametric local polynomial regression with the polynomial degree set to one. Finally, although point estimation is done with (local weighted) linear models, we stress that inference is then conducted using robust bias correction, which we adapt for heterogeneity below.

\cite{Calonico-Cattaneo-Farrell-Titiunik_2019_RESTAT} studied the addition of the pretreatment covariates to this model, but aiming exclusively for more efficient estimation of the RD average treatment effect $\tau$, not for heterogeneity analysis. They showed that efficiency gains could be achieved by including covariates linearly, but \emph{not} interacting them with the treatment or running variable. Specifically, for pretreatment covariates $\bZ_i$ (used for efficiency only and hence necessarily distinct from $\bW_i$), \cite{Calonico-Cattaneo-Farrell-Titiunik_2019_RESTAT} studied $\widetilde{\tau}$ from the (local weighted) least squares fit given by
\begin{equation}
	\label{eqn:reg 2}
	\widetilde{Y}_i = \widetilde{\mu}  + \widetilde{\tau} T_i + \widetilde{\omega}_1 X_i + \widetilde{\omega}_2 T_i X_i + \widetilde{\bgamma}'\bZ_i.
\end{equation}
Unlike $\dot{\tau}$, the estimated coefficient $\widetilde{\tau}$ cannot be obtained from two separate fits because of the common coefficient on $\bZ_i$. Despite being linear in both the running variable and covariates, equation \eqref{eqn:reg 2} is nonparametric in $X_i$ (local polynomial of degree one, localized using the bandwidth $h$) but it is a linear projection in $\bZ_i$, and thus the interpretation of the estimated coefficient $\widetilde{\bgamma}$ is only as a best linear projection (unless the population mean function happens to be linear). That is, the nonparametric smoothing is only on the running variable. This respects the local nature of the RD design while being simple and tractable in applications. \cite{Calonico-Cattaneo-Farrell-Titiunik_2019_RESTAT} proved that $\widetilde{\tau}$ recovers the average effect $\tau$, characterized conditions for efficiency gains, and provided optimal bandwidth selection and valid inference using bias robust bias correction. \cite{Ma_Yu2024_WP} strengthened the efficiency argument for this regression fit. The exclusion of interactions with the covariates in \eqref{eqn:reg 2} is crucial for correct estimation of the average effect $\tau$ of \eqref{eqn:ate}, but by construction does not allow for learning heterogeneity. Because our present goal is explicitly to learn about heterogeneous treatment effects, we need to allow some form of interaction between the pretreatment covariates and treatment indicator. 

Ideally, one might try to recover the (local to $X_i=0$) conditional average treatment effect (CATE) function fully flexibly in $\bW_i$, which is defined by
\begin{equation}
    \label{eqn:cate}
    \kappa(\bw) = \E[Y_i(1) - Y_i(0) | X_i = 0, \bW_i = \bw].
\end{equation}
Absent further assumptions, $\kappa(\bw)$ is the difference between two $(1+d)$ dimensional nonparametric regressions evaluated at the point $(0,\bw)\in\mathbb{R}^{1+d}$. While such a setting is possible to handle theoretically, it is a challenging problem in real-world settings due to the curse of dimensionality and the need to choose many tuning parameters. Following empirical practice, we aim for something more tractable. 

Indeed, one possible explanation for the popularity of RD in empirical work is that it combines a credible causal identification design with perhaps the simplest nonparametric problem, namely \emph{one}-dimensional nonparametric regression at a single point, $X_i=0$. This is reflected in the simplicity of \eqref{eqn:reg 1} and \eqref{eqn:reg 2}. Thus, rather than designing a fully general procedure for obtaining $\kappa(\bw)$, we focus on specifications that share the same practical appeal. 

We consider the local weighted least squares regression with full interactions between the treatment indicator and the pretreatment covariates, which may be discrete or continuous. Our main task is analyzing the interpretation and properties of the regression fit
\begin{equation}
	\label{eqn:reg 3}
	\widehat{Y}_i = \ahat + \widehat{\theta} T_i 
                  + \widehat{\boldsymbol{\lambda}}'\bW_i + \widehat{\boldsymbol{\xi}}' T_i \bW_i
                  + \widehat{\omega}_1 X_i + \widehat{\omega}_2 T_i X_i
                  + \widehat{\boldsymbol{\omega}}_3' X_i \bW_i + \widehat{\boldsymbol{\omega}}_4'T_i X _i \bW_i,
\end{equation}
and of the corresponding estimate of the CATE function $\kappa(\bw)$ at the cutoff given by
\begin{equation}
    \label{eqn:cate hat}
    \widehat{\kappa}(\bw) = \thetahat + \widehat{\boldsymbol{\xi}}'\bw.
\end{equation}
This regression is the natural next step in the progression of \eqref{eqn:reg 1} and \eqref{eqn:reg 2}. However, the interpretation of the coefficients changes: the coefficient on $T_i$ no longer recovers the average effect $\tau$ but rather $\kappa(\bOO)$, the CATE for the baseline group $\bw=\bOO$ (where $\bOO$ is a vector of zeroes), under appropriate identifying assumptions (Assumption \ref{asmpt:linear} below). Moreover, while the coefficients on $\bZ_i$ in \eqref{eqn:reg 2} are largely uninterpretable since they are for efficiency purposes only, here the coefficients $\widehat{\boldsymbol{\xi}}$ give the difference from the baseline group, either as the discrete change for binary covariates or, under our functional coefficient assumption below, as the slope of $\kappa(\bw)$ with respect to continuous $\bw$. Although we omit it from our main discussion, the supplemental appendix provides results for estimation and inference on derivatives of $\kappa(\bw)$, covering kink RD designs and related settings \citep{Card-Lee-Pei-Weber_2015_ECMA}.

As with Equations \eqref{eqn:reg 1} and \eqref{eqn:reg 2}, the fit \eqref{eqn:reg 3} is a local polynomial (of degree 1) approximation in $X_i$, with bandwidth $h$ and kernel weights $K(X_i/h)$ around the cutoff $c=0$. In this case, this is true for the main effects as well as the interaction terms: while $\widetilde{\bgamma}$ in \eqref{eqn:reg 2} is a linear projection coefficient with no intended approximation power (or causal interpretation), the semiparametric fit \eqref{eqn:reg 3} delivers local polynomial approximations to unknown functions of $X_i$, on either side of the cutoff, that form linear interactions with $\bW_i$. This is formalized in Assumption \ref{asmpt:linear} below and in the supplemental appendix in full generality (see Section SA1.3). Importantly, such an approximation cannot consistently estimate an arbitrary $\kappa(\bw)$; its causal interpretation crucially relies on Assumption \ref{asmpt:linear}. This is true because the regression \eqref{eqn:reg 3} is nonparametric in $X_i$ but not in $\bW_i$; there is no approximation to unknown functions of $\bW_i$. 

In applications, the vector $\bW_i$ must therefore be specified by the researcher to capture the relevant heterogeneity. Changing the covariates included in the regression, or their specification, changes the target estimand and its interpretation. With discrete variables, the vector $\bW_i$ should contain indicators for every relevant category, unless the researcher decides ex-ante that certain categories are not to be studied. For example, if two binary variables $(B_{i,1},B_{i,2})$ are present, then $\bW_i = \big( (1-B_{i,1})B_{i,2},\  B_{i,1}(1-B_{i,2}),\ B_{i,1}B_{i,2}\big)'$, with $B_{i,1}=B_{i,2}=0$ being the baseline category. A multi-level discrete variable would require that $\bW_i$ contain an indicator for each level beyond the baseline, which forms a collection of orthogonal binary variables. For example, if $\mathcal{W}$ is the finite set of values taken by the covariate of interest $W_i$ (e.g., years of education), then $\bW_i=(\indic(W_i=w), w\in\mathcal{W})'$. In all of these cases, the structure of Assumption \ref{asmpt:linear} holds without loss of generality. Continuous variables can be included directly into $\bW_i$, in which case $\widehat{\boldsymbol{\xi}}$ has the standard interpretation of a slope coefficient in a linear model. Further, transformations of continuous variables (such as polynomials) can be added for additional flexibility. Alternatively, the researcher might estimate the heterogeneous treatment effects one variable at a time, which may not capture the full heterogeneity pattern but may provide a simple and intuitive summary of different dimensions of heterogeneity. Section \ref{sec:application} illustrates different approaches and interpretations empirically. (See \cite{Calonico-Cattaneo-Farrell-Palomba-Titiunik_2025_Stata} for further examples and discussion.)

As with \eqref{eqn:reg 1}, and in contrast to \eqref{eqn:reg 2}, the coefficients in \eqref{eqn:reg 3} can be obtained by first fitting separate regressions on either side of the cutoff, and then taking appropriate differences. It is also possible to run separate models for different subgroups of discrete variables via \eqref{eqn:reg 1} to recover the estimates in \eqref{eqn:reg 3}, but the joint estimation procedure is simple and intuitive and thus appealing for practice. Joint estimation is natural if additional controls are added: following \cite{Calonico-Cattaneo-Farrell-Titiunik_2019_RESTAT}, it is possible to include some covariates for efficiency gains, while other covariates for heterogeneity analysis, thereby obtaining more efficient estimation and inference for $\kappa(\bw)$. This mixed covariate adjustment approach augments \eqref{eqn:reg 3} with additional covariates $\bZ_i$ that are not interacted with $T_i$ nor $X_i$, but are interacted with $\bW_i$. That is, one would fit
\begin{equation}
   \label{eqn:reg 4}
   \check{Y}_i = \check{\alpha} + \check{\theta} T_i 
               + \check{\boldsymbol{\lambda}}'\bW_i + \check{\boldsymbol{\xi}}' T_i \bW_i
               + \check{\omega}_1 X_i + \check{\omega}_2 T_i X_i
               + \check{\boldsymbol{\omega}}_3' X_i \bW_i + \check{\boldsymbol{\omega}}_4'T_i X _i \bW_i
               + \check{\bgamma}_1'\bZ_i
               +\check{\bgamma}_2'(\bZ_i\otimes\bW_i).
\end{equation}
When $\bW_i$ are binary, this approach is theoretically justified by the results presented in the upcoming sections coupled with the theory of \cite{Calonico-Cattaneo-Farrell-Titiunik_2019_RESTAT}. For the general case, the results in the supplemental appendix can be extended to accommodate the fit in \eqref{eqn:reg 4}, though the restriction on the coefficients of $\bZ_i$ would be slightly different than for category-wise estimation, and additional regularity conditions would be needed.

Unlike Equation \eqref{eqn:reg 2}, our proposed fit \eqref{eqn:reg 3} does not yield an estimator of the average treatment effect \eqref{eqn:ate}. This is implicit in Theorem \ref{thm:plim} below, and was already shown by \cite{Calonico-Cattaneo-Farrell-Titiunik_2019_RESTAT}. There are two potential remedies to this, neither of which we recommended in practice. First, the covariates can be demeaned on each side of the cutoff. This would follow the logic for linear regression adjustments in randomized experiments, but in this case, the demeaning would amount to Nadaraya–Watson estimation at a boundary point with $\bW_i$ as the outcome, thereby hampering the entire RD estimation due to the presence of smoothing bias. Second, one could average the fitted CATEs, but this again essentially amounts to Nadaraya–Watson estimation at a boundary point, now using $\widehat{\kappa}(\bw)$ as the outcome. In both cases, the convergence rate will be slow because the misspecification bias is of order $h$, and is thus impractical for inference (e.g., severe undersmoothing would be needed). Therefore, in applications, researchers should only use our proposed methods for heterogeneity analysis and rely on the established methods underlying \eqref{eqn:reg 1} or \eqref{eqn:reg 2} to estimate the average RD treatment effect.

Finally, a related motivation for the specification \eqref{eqn:reg 4} is to control for panel or group structures. It is common in some RD applications to include fixed effects for individuals, groups, or time periods. Identification, estimation, and inference then follow from the same arguments presented in the upcoming sections. In this way, our proposed RD methodology, although developed for the cross-sectional setting, can be easily applied to panel data, provided weighted linear least squares methods with additive or interactive fixed effects are used.

\section{Main Results}
    \label{sec:theory}

We turn to the theoretical properties of the regression fit in \eqref{eqn:reg 3}, and particularly the CATE estimator \eqref{eqn:cate hat}. We provide a complete set of practical tools for identification, estimation, and inference. The following standard regularity conditions are imposed. Recall that the cutoff is $c=0$ without loss of generality, and let $x_{\mathtt{L}} < 0 < x_{\mathtt{U}}$. All omitted details are given in the supplemental appendix (see Section SA1.3).

\begin{assumption}[Sharp RD Design]\label{asmpt:main}
    The data $(Y_i, X_i, \bW_i'), i=1,\ldots, n$, where $T_i=\indic\{X_i \geq 0\}$ and $Y_i = T_i Y_i(1) + (1-T_i) Y_i(0)$, are a random sample obeying the following:
    \begin{enumerate}[label=\emph{(\alph*)}]
        \item $X_i$ admits a continuous, bounded, and bounded away from zero Lebesgue density; 
        \item For $t\in \{0,1\}$, $\bw\in\R^d$, and all $x\in [x_{\mathtt{L}},x_{\mathtt{U}}]$: $\V[Y_i(t)|X_i=x]$, $\E[|Y_i(t)|^4|X_i=x, \bW_i=\bw]$, $\E[\bW_i|X_i=x]$, $\E[\bW_i\bW_i'|X_i=x]$, $\E[|\bW_i|^4|X_i=x]$, $\E[\bW_i \V[Y_i(t)|X_i,\bW_i]| X_i=x]$, and $\E[\bW_i\bW_i'\V[Y_i(t)|X_i,\bW_i]| X_i=x]$ are continuous in $x$;
        \item $\E[K(X_i/h) \bA_i \bA_i']$ is invertible, where $\bA_i = (1,X_i,X_i^2)' \otimes (1, T_i,\bW_i', T_i \bW_i')'$; and
        \item $K(\cdot)$ is nonnegative and continuous on its support $[-1,1]$.
    \end{enumerate}

\end{assumption}

These assumptions are on par with prior RD literature, only adding minimal extra regularity to accommodate the covariates $\bW_i$ for heterogeneity analysis via interactions. The most important substantive condition we require is that $\bW_i$ are pretreatment. In particular, continuity of $\E[\bW_i|X_i=x]$ means there is no RD treatment effect on the covariates, while the same is required for higher moments for the asymptotic analyses. Most commonly, this holds when $\bW_i$ is determined before treatment is assigned. Parts (b) and (c) are regularity conditions needed for valid estimation and (robust bias-corrected) inference. Part (c) is a familiar rank condition for least squares regression, adapted to \eqref{eqn:reg 3} and its bias-corrected counterpart. The regularity conditions are slightly stronger than those imposed by \cite{Calonico-Cattaneo-Farrell-Titiunik_2019_RESTAT} because of the more flexible regression required for recovery of heterogeneous treatment effects, $\kappa(\bw)$, compared to the average treatment effect $\tau$ considered in that paper (compare \eqref{eqn:reg 2} to \eqref{eqn:reg 3}). Part (d) is a standard assumption when employing kernel-weighted nonparametric regressions.

To characterize exactly when the CATE function can be recovered based on a local least squares fit with interactions, we introduce the following identifying assumption.

\begin{assumption}[Local CATE Structure]\label{asmpt:linear}
    For $\bw\in\R^d$, and all $x\in [x_{\mathtt{L}},x_{\mathtt{U}}]$:
    \begin{align*}
        \E[Y_i(1)-Y_i(0) | X_i=x, \bW_i=\bw] = \theta(x) + \bxi(x)'\bw
     \end{align*}
    and
    \begin{align*}
        \E[Y_i(0) | X_i=x, \bW_i=\bw] = \alpha(x) + \blambda(x)'\bw,
    \end{align*}
    where the functions $\theta(x)$, $\bxi(x)$, $\alpha(x)$, and $\blambda(x)$ are thrice continuously differentiable. 
\end{assumption}

This assumption requires the potential outcomes to obey a local semilinear-in-$\bW$-parameters model, that is, a local linear functional coefficient model, which naturally aligns with the estimation procedure in \eqref{eqn:reg 3}. It follows from this assumption that the CATE function (at the cutoff $x=c=0$) is linear in $\bw$:
\begin{align*}
    \kappa(\bw) = \Big.\E[Y_i(1)-Y_i(0) | X_i=0, \bW_i=\bw]  = \theta(0) + \bxi(0)'\bw.
\end{align*}
From an econometric point of view, Assumption \ref{asmpt:linear} means that estimation of the CATE function remains a one-dimensional nonparametric problem and, further, that the bias can be removed up to the same order as learning the average treatment effect in standard sharp RD settings. This is important for maintaining the empirical tractability of RD analyses. When $\bW_i$ contains continuous variables, the linearity (in parameters) places a restriction on the ``long'' regression functions $\E[Y_i(t) | X_i=x, \bW_i=\bw]$, and thus caution must be taken when interpreting the results. Note that the assumption requires linearity in parameters, and so transformations, such as polynomials, can be included in $\bW_i$ to weaken the implied identifying restrictions. Alternatively, when the covariates are binary, $\bW_i\in\{0,1\}^d$, Assumption \ref{asmpt:linear} is automatically satisfied. Thus, our proposed joint estimation and inference methods are automatically valid for causal heterogeneous RD treatment effects based on discrete covariates (via the construction of orthogonal binary variables for each category), thereby offering a formal causal interpretation for the covariate-interacted local least squares estimates in RD designs.

More generally, Assumption \ref{asmpt:linear} gives a causal interpretation to the commonly used estimation approach in \eqref{eqn:reg 3}, but Assumption \ref{asmpt:main} would be sufficient for all our estimation and inference results if researchers were interested in studying a best mean square approximation of the mean functions $\E[Y_i(t) | X_i = c, \bW_i = \bw]$---that is, if the parameter of interest were simply the probability limit of the estimated coefficients in \eqref{eqn:reg 3}. However, those are often neither useful nor interpretable as heterogeneous causal effects. For more details, see the supplemental appendix.

We summarize our first identification and estimation result in the following theorem. Let $\to_\P$ denote convergence in probability as $n\to\infty$.
\begin{thm}\label{thm:plim}
    Suppose Assumptions \ref{asmpt:main} and \ref{asmpt:linear} hold, $h\to0$, and $nh\to\infty$. Then, $\widehat{\kappa}(\bw) \to_\P \kappa(\bw)$.
\end{thm}
This result establishes that the regression fit in \eqref{eqn:reg 3} correctly captures the heterogeneous causal effects in RD designs. Without Assumption \ref{asmpt:linear}, the probability limit of the coefficients in \eqref{eqn:reg 3} do not appear to have a causal interpretation or useful closed-form expressions, and $\widehat{\kappa}(\bw)$ does not recover a useful object (see the supplemental appendix). This is in contrast to recovering the average RD effect, $\tau$, using $\widetilde{\tau}$ from \eqref{eqn:reg 2}, where further restrictions on the underlying data generating process are not needed despite the inclusion of pretreatment covariates (for efficiency purposes). 

\subsection{Mean Squared Error and Bandwidth Selection}
    \label{sec:bw}

Estimation and inference rely crucially on selecting a bandwidth $h$ that appropriately localizes to the cutoff $c = 0$. This problem is well studied in the methodological literature on RD designs \citep{Arai-Ichimura_2018_QE,Calonico-Cattaneo-Farrell_2020_ECTJ}. The choice of kernel weights is typically less consequential, and typically a linear fit is used for point estimation \cite[see][for more discussion]{Pei-Lee-Card-Weber_2021_JBES}. We develop a bandwidth selection method that is optimal in a mean squared error (MSE) sense for estimation of heterogeneous RD treatment effects. Such bandwidth yields optimal point estimates of heterogeneous RD treatment effects, making it an appropriate choice for reporting point estimates in empirical applications. However, using such a bandwidth yields invalid inference in general, and thus, in the next section, we develop inference methods using the standard approach of robust bias correction \citep{Calonico-Cattaneo-Titiunik_2014_ECMA}. 

To characterize the optimal bandwidth for treatment effect estimation, we establish an MSE expansion of the coefficients $\widehat{\boldsymbol{\varsigma}} = (\widehat{\theta},\widehat{\boldsymbol{\xi}}')'$ in the local regression \eqref{eqn:reg 3} around their population counterparts. The probability limit of this vector is, under Assumption \ref{asmpt:linear}, $\boldsymbol{\varsigma} = (\theta(0), \boldsymbol{\xi}(0)')'$. Linear combinations of elements of $\boldsymbol{\varsigma}$ yield the CATE function for different groups or contrasts. Let $\select$ be the conformable vector that selects or combines the appropriate elements. For example, in the case of a binary $W_i$ ($d=1$), the base category subgroup can be studied as $\widehat{\theta} = \select'\widehat{\boldsymbol{\varsigma}}$ with $\select = (1,0)'$, whereas testing group differences would correspond to $\widehat{\boldsymbol{\xi}} = \select'\widehat{\boldsymbol{\varsigma}}$ with $\select = (0,1)'$. In general, setting $\select = (1,\bw)'$ for a given $\bw$ yields $\select'\widehat{\boldsymbol{\varsigma}} = \thetahat + \widehat{\boldsymbol{\xi}}'\bw = \widehat{\kappa}(\bw)$, recovering the CATE estimator of \eqref{eqn:cate hat}. 

We now discuss the bandwidth selection result. The general MSE expansion is more cumbersome notationally; we provide it in the supplemental appendix (Section SA2.6). As is customary, after some approximations, it depends on the leading variance and squared bias: $\mathsf{Var}[\select'\widehat{\boldsymbol{\varsigma}}] = \frac{1}{n h} \mathsf{V}_{\select}$ and $\mathsf{Bias}^2[\select'\widehat{\boldsymbol{\varsigma}}] = h^4 \mathsf{B}_{\select}^2$, respectively. The rates reflect the specific structure of the estimator in \eqref{eqn:reg 3}, a local linear approximation based only on the scalar $X_i$. The constants $\mathsf{V}_{\select}$ and $\mathsf{B}_{\select}$ also capture the specific features of the estimator and target estimand (hence they are functions of $\select$). The following theorem summarizes our bandwidth selection result.

\begin{thm}[MSE-optimal Bandwidth]
    \label{thm:mse}
    Suppose Assumptions \ref{asmpt:main} and \ref{asmpt:linear} hold, and $h\to0$ and $nh^3\to\infty$. Then, provided that $\mathsf{B}_{\select} \neq 0$, the MSE-optimal bandwidth for estimating $\select'\widehat{\boldsymbol{\varsigma}}$ is
    \begin{align}\label{eqn:h}
        h^\star_{\select} = \Big( \frac{\mathsf{V}_{\select} }{4\mathsf{B}_{\select}^2}\frac{1}{n} \Big)^{1/5},
    \end{align}
    where $\mathsf{V}_{\select}$ and $\mathsf{B}_{\select}$ are given in the supplemental appendix (Section SA2.6).
\end{thm}

While the optimal bandwidth depends on the specific target estimand through the constant terms, its convergence rate is not affected because equation \eqref{eqn:reg 3} is still a one-dimensional nonparametric (in $X_i$) estimation problem. Again, consider the leading case of a binary covariate. If one group is substantially smaller than the other, then, all else equal, the variance will be higher for that group. It could also be that the two groups have conditional expectations with different local curvature, resulting in different biases. Thus, ideally, the bandwidth should be different for each estimand. However, this is not necessarily practical, and the specific constant used in bandwidth selection is often less important empirically than ensuring that $h$ is roughly of the right magnitude. In the context of heterogeneity analysis in RD designs, this perspective supports the common practice of employing the same bandwidth for the estimation of all heterogeneous RD treatment effects. Section \ref{sec:application} provides further discussion in the context of our empirical application.  

Our result offers clear practical guidance and justification for several approaches. Researchers may decide that different groups should have substantially different bandwidths. Alternatively, they may prefer to conduct all analyses using a common bandwidth. This is easily accommodated by the joint fitting in \eqref{eqn:reg 3}. This common bandwidth could be selected as the same bandwidth used for the average effect in \eqref{eqn:reg 1}. One could also use the median of all the relevant bandwidths or, for robust inference, the smallest of the set. Using different bandwidths for different tasks may yield more accurate inference in some applications.

\subsection{Robust Bias-Corrected Inference}
    \label{sec:inference}

We now state our main results for inference on heterogeneous RD treatment effects. We consider the case of estimating the joint regression \eqref{eqn:reg 3} using an MSE-optimal bandwidth for point estimation and then conducting inference using robust bias correction---the supplemental appendix gives more general theoretical results (see Section SA2.6). Our main inference procedures rely on a Gaussian distributional approximation for the robust bias corrected versions of $\widehat{\boldsymbol{\varsigma}} = (\widehat{\theta},\widehat{\boldsymbol{\xi}}')'$ of the local regression \eqref{eqn:reg 3}, along with valid standard errors. This can then be specialized to different use cases depending on which coefficients are combined or contrasted via the continuous mapping theorem. 

Using an MSE-optimal bandwidth means that the limiting distribution of $\widehat{\boldsymbol{\varsigma}}$ is not centered at $\boldsymbol{\varsigma} = (\theta(0), \boldsymbol{\xi}(0)')'$, under Assumption \ref{asmpt:linear}, a fact that renders inference invalid. More generally, it is shown in the supplemental appendix that
\begin{align*}
    \sqrt{nh} \big( \widehat{\boldsymbol{\varsigma}} - \boldsymbol{\varsigma} - h^2 \mathsf{B} \big) \rightsquigarrow \mathsf{Normal}(0,\mathsf{V}),
\end{align*}
where $\rightsquigarrow$ denotes convergence in distribution as $h\to0$ and $nh\to\infty$, and $\mathsf{B}$ and $\mathsf{V}$ the leading asymptotic bias and asymptotic variance of the estimator $\widehat{\boldsymbol{\varsigma}}$. In particular, $\mathsf{V}_{\select} = \select'\mathsf{V}\select$ and $\mathsf{B}_{\select} = \select'\mathsf{B}$. It thus follows that a choice of bandwidth proportional to $n^{-1/5}$ (e.g., $h^\star_{\select}$ developed above) leads to a MSE-optimal point estimator $\widehat{\boldsymbol{\varsigma}}$ but invalid inference due to a first-order bias in the distributional approximation. 

For inference in RD applications, robust bias correction is a popular way to deal with this problem. In brief, there are two key ingredients to robust bias correction: (i) an estimate of $h^2 \mathsf{B}$ is subtracted from the point estimate $\widehat{\boldsymbol{\varsigma}}$, which amounts to bias correction, and (ii) the standard errors are adjusted to account for the additional estimation error of the bias correction step, which amounts to variance correction. Robust bias-correction continues to rely on a Gaussian distributional approximation, but inference is based on the corrected statistics 
\begin{equation}
    \label{eqn:rbc}
    \widehat{\boldsymbol{\varsigma}}_{\rm rbc} = \widehat{\boldsymbol{\varsigma}} - h^2 \widehat{\mathsf{B}}
    \qquad\text{and}\qquad
    \widehat{\mathsf{V}}_{\rm rbc} = \widehat{\rm var}(\widehat{\boldsymbol{\varsigma}}_{\rm rbc}),
\end{equation}
where $\widehat{\mathsf{B}}$ is an estimator of the bias of $\widehat{\boldsymbol{\varsigma}}$, and $\widehat{\mathsf{V}}_{\rm rbc}$ is an estimator of the variance of $\widehat{\boldsymbol{\varsigma}}_{\rm rbc}$.

The robust bias correction method for RD design has been discussed in detail, see \cite{Cattaneo-Idrobo-Titiunik2019_book}, and references therein. See also the supplemental appendix for omitted technical details, including results on heteroskedastic- and cluster-robust standard errors \citep{Cameron2015_JHR,mackinnon2023ClusterrobustInferenceGuide}. Robust bias correction is known to have excellent theoretical and numerical properties \citep{Calonico-Cattaneo-Farrell_2018_JASA,Calonico-Cattaneo-Farrell_2022_Bernoulli}, and has been shown to yield correct inference in extensive validation empirical studies \citep{Hyytinen-Tukiainen-etal2018_QE,DeMagalhaes-etal_2025_PA}.

We obtain the following asymptotic normality result, based on the estimates and standard errors in \eqref{eqn:rbc}, for any specific estimand defined by $\select$. The exact formulas for $\widehat{\mathsf{B}}$ and $\widehat{\mathsf{V}}_{\rm rbc}$ are given in the supplemental appendix (Section SA2.2).

\begin{thm}[Asymptotic Normality]\label{thm:clt-rbc}
    Suppose Assumptions \ref{asmpt:main} and \ref{asmpt:linear} hold, and $h=h^\star_{\select}$. Then, 
    \begin{align*}
        \frac{\select'\widehat{\boldsymbol{\varsigma}}_{\rm rbc} - \select'\boldsymbol{\varsigma}}{\sqrt{\select'\widehat{\mathsf{V}}_{\rm rbc}\select}} \rightsquigarrow \mathsf{Normal}(0,1).
    \end{align*}
\end{thm}

The supplemental appendix gives a more general result, which includes a Gaussian distributional approximation without Assumption \ref{asmpt:linear} (and hence with a different centering). Theorem \ref{thm:clt-rbc} can be used to carry out valid inference for any of the estimands we have discussed, including the CATE function itself and comparisons between different values of $\bw$, for discrete or continuous $\bW_i$. For example, an approximate robust bias-corrected 95\% confidence interval for $\kappa(\bw)$ based on $\select = (1, \bw)'$ is
\begin{align*}
    \Big[\;\widehat{\kappa}_{\rm rbc}(\bw) - 1.96 \sqrt{\select'\widehat{\mathsf{V}}_{\rm rbc}\select}
         \;,\;
         \widehat{\kappa}_{\rm rbc}(\bw) + 1.96 \sqrt{\select'\widehat{\mathsf{V}}_{\rm rbc}\select}\;\Big].
\end{align*}
Robust hypothesis tests and $p$-values can also be constructed in the standard manner. For example, assuming $\bW_i\in\{0,1\}$, a hypothesis test of no heterogeneous RD treatment effect boils down to testing $\mathsf{H}_0:\xi(0) = 0$, which corresponds to employing $\select=(0,1)'$. Such inference procedures should be reported along with the point estimates derived from $\widehat{\boldsymbol{\varsigma}}$ (as opposed to reporting $\widehat{\boldsymbol{\varsigma}}_{\rm rbc}$). 

\section{Application}
    \label{sec:application}

To illustrate our methodological results, we revisit the analysis by \cite{Akhtari-Moreira-Trucco2022_AER}, which examines the impact of political turnover in Brazilian mayoral elections on local public service provision using an RD design based on close elections. \cite{KlasnjaTitiunik2017-APSR} first studied these data and RD design, and documented large negative RD effects of party incumbency based on Brazilian mayoral elections, comparing municipalities where the incumbent party barely loses (resulting in political turnover) to those where the incumbent party barely wins (no turnover): a bare victory results in a large reduction in the probability of victory in the following election, which they attribute to deficient accountability resulting from weak parties and term limits. 

Using a similar RD design based on close elections, \cite{Akhtari-Moreira-Trucco2022_AER} find that political turnover leads to a sudden expansion of municipal bureaucracy, with new personnel appointed across multiple sectors at both managerial and non-managerial levels. They also investigate educational outcomes, reporting increased replacement rates among school personnel in municipally controlled schools, along with declines in student test scores. We focus on heterogeneous effects on headmaster replacement (i.e., whether a headmaster is new to the school) by municipality income (we rescale income into hundreds to improve the numerical presentation in Table \ref{table:AMT} below). See their Table A.21.

The analysis of \cite{Akhtari-Moreira-Trucco2022_AER} highlights two key methodological difficulties researchers have faced when studying heterogeneous treatment effects in RD without methodological and theoretical guidance in the literature, as discussed above. First, \cite{Akhtari-Moreira-Trucco2022_AER} use an optimal data-driven bandwidth selection method \citep{Calonico-Cattaneo-Farrell-Titiunik_2019_RESTAT} to localize at the cutoff. However, both estimation and inference rely solely on local linear fitting. Although estimation using linear regression (specifically, their implementation of \eqref{eqn:reg 2}) is correct, valid inference requires robust bias correction, as standard errors from the linear fit are invalid when paired with the optimal bandwidth. In our replication, we revisit their heterogeneity analysis using robust bias-corrected inference. Our results, presented in Table \ref{table:AMT}, report confidence intervals instead of standard errors, following best practices for RD analysis.

The second key aspect is that the heterogeneity variable of interest, income, is continuous. \cite{Akhtari-Moreira-Trucco2022_AER} binarize it to compare municipalities above and below the median income level. We first replicate this exercise using our methods. The results are shown in the top two rows of Panel (b), Table \ref{table:AMT}. Our point estimates are nearly identical to theirs (0.373 and 0.122 compared 0.389 and 0.126 in their Table A.21), where the differences are due to slight bandwidth variations. Next, we incorporate valid uncertainty measures using robust bias correction, revealing that the treatment effect is statistically significant only for low (below-median) income municipalities. We also formally test the null hypothesis that the treatment effect is the same across income groups, finding strong evidence against it: our robust inference yields a 95\% confidence interval of [-0.422, -0.078] for the difference, which excludes zero and reveals statistically significant heterogeneity. We also extend this discretized analysis by studying treatment effects by income quartiles and deciles, rather than a binary median split. This refinement adds nuance to the original study, showing that variation in treatment effects is more pronounced across the median than within the upper or lower groups. These results further strengthen the original findings.

\begin{figure}
	\caption{Graphical Presentation of Overall and Heterogeneous RD Treatment Effects by Income}
	\label{fig:AMT}
	\centering
	\begin{subfigure}[b]{0.32\textwidth}
		\centering
		\includegraphics[trim={0cm 0cm 1cm 1cm}, clip, width=1.2\textwidth]{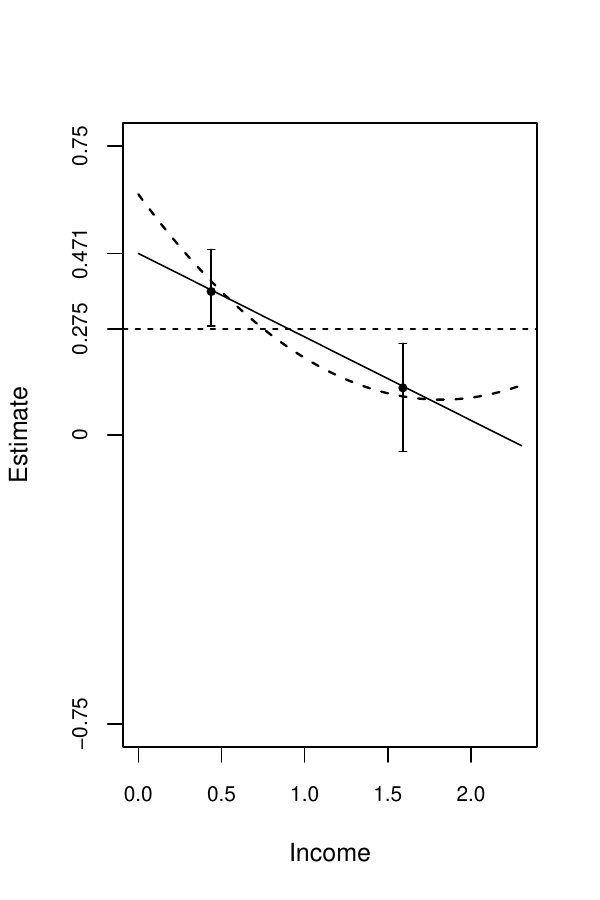}
		\caption{Median}
		\label{fig:AMTa}
	\end{subfigure}
	\begin{subfigure}[b]{0.32\textwidth}
		\centering
		\includegraphics[trim={0cm 0cm 1cm 1cm}, clip, width=1.2\textwidth]{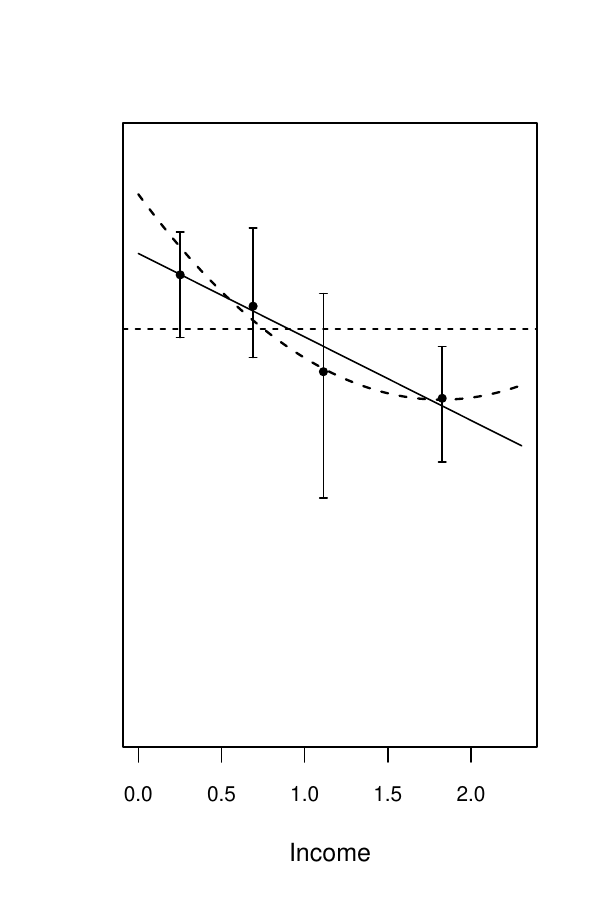}
		\caption{Quartiles}
		\label{fig:AMTb}
	\end{subfigure}
	\begin{subfigure}[b]{0.32\textwidth}
		\centering
		\includegraphics[trim={0cm 0cm 1cm 1cm}, clip, width=1.2\textwidth]{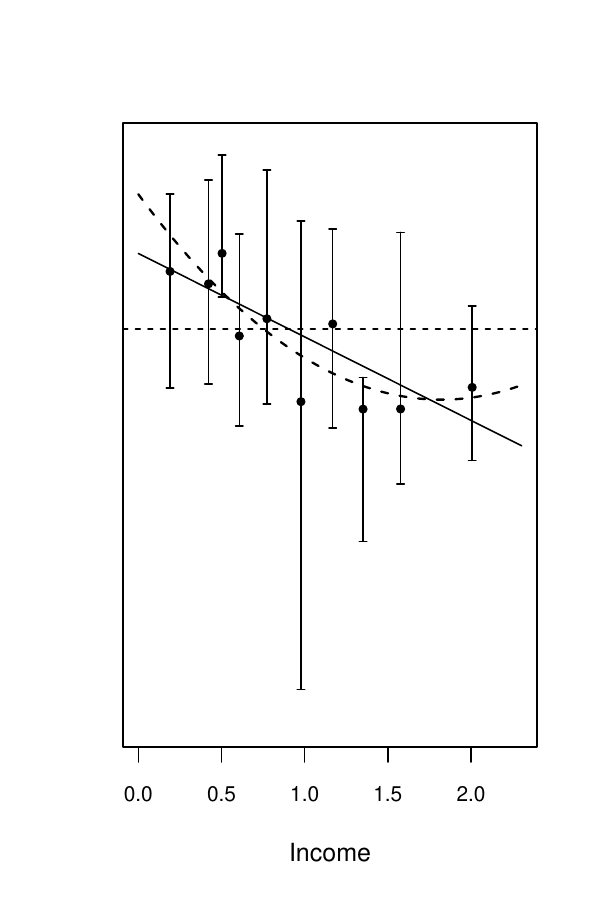}
		\caption{Deciles}
		\label{fig:AMTc}
	\end{subfigure}
    
    \scriptsize{
    \begin{flushleft}Notes:\end{flushleft}\vspace{-0.2in}
    \begin{enumerate}[label=\normalfont(\roman*),noitemsep,leftmargin=*]
        \item The horizontal dashed line is the estimated overall RD treatment effect: $\dot{\tau}$ of \eqref{eqn:reg 1}. See Table \ref{table:AMT}, Panel (a), for the precise empirical result plotted.
        
        \item The dots show the heterogeneous treatment effects for discretized income, with vertical bars representing their associated robust bias-corrected $95\%$ confidence intervals. Subfigure (a) reports discretized heterogeneity below and above median, subfigure (b) reports discretized heterogeneity by quartiles, and subfigure (c) reports discretized heterogeneity by deciles. See Table \ref{table:AMT}, Panel (b), for the precise empirical results plotted.
        
        \item The solid line shows the estimated semi-linear heterogeneous treatment effect using income as a continuous variable linearly: $\widehat{\kappa}({\tt income}_i) = \widehat{\theta} + \widehat{\xi} \cdot {\tt income}_i$. See Table \ref{table:AMT}, Panel (c), for the precise intercept and slope estimates.

        \item The nonlinear dashed line shows the estimated semi-linear heterogeneous treatment effect using income as a continuous variable quadratically: $\widehat{\kappa}({\tt income}_i) = \widehat{\theta} + \widehat{\xi}_1 \cdot {\tt income}_i + \widehat{\xi}_2 \cdot {\tt income}_i^2$. See Table \ref{table:AMT}, Panel (d), for the precise estimates.

        \item See the footnote of Table \ref{table:AMT}, and the replication files at \url{https://rdpackages.github.io/replication/}, for more details.

    \end{enumerate}
    }
\end{figure}

Finally, we analyze income directly, as a continuous variable. The specification remains \eqref{eqn:reg 3}, but its interpretation is now governed by Assumption \ref{asmpt:linear} with an intercept and slope coefficient. In Table \ref{table:AMT}, Panel (c) reports estimated intercept and slope based on a linear-in-income model with $\bW_i=\mathtt{income}_i$, while Panel (d) considers a quadratic-in-income model with $\bW_i=(\mathtt{income}_i,\mathtt{income}_i^2)$. Comparing the discretized heterogeneity with these linear and quadratic models, as shown graphically in Figure \ref{fig:AMT}, reveals that the two continuous models effectively capture the overall heterogeneity in this application. Notably, the continuous specifications highlight differences in income distribution within each group, particularly emphasizing that the top quartile has a much wider range than the others---an important finding with policy implications.

Ultimately, the decision of how to best capture heterogeneity is up to the researcher. The discretized and continuous versions have their strengths and limitations. The continuous approach, particularly when compared to a fine discretization, may yield substantial precision improvements, but requires imposing the structure of Assumption \ref{asmpt:linear}. The choice ultimately depends on the specific application and the researcher's objectives. Our framework provides rigorous estimation and inference methods for both cases.

\section{Conclusion}
    \label{sec:conclusion}

We have shown how treatment effect heterogeneity can be effectively analyzed in RD settings using a tractable and transparent linear-interactions model. Our approach avoids the complexities of high-dimensional nonparametric regression while seamlessly integrating into standard RD practice. By incorporating principled bandwidth selection and robust bias-corrected inference, we aim to bring the same level of rigor and standardization to heterogeneous effects as currently exists for average treatment effects. Our analysis has focused exclusively on sharp RD designs, including kink RD designs in the supplemental appendix. Extending these methods to other settings—particularly fuzzy designs—remains an important direction for future research.

\clearpage

\begin{table}
\renewcommand{\arraystretch}{1}
    \caption{Overall and Heterogeneous RD Treatment Effects by Income: Political Turnover and Headmaster Replacement in Municipalities.}
    \label{table:AMT}
    \centering
    \resizebox{0.88\textwidth}{!}{
    \begin{tabular}{lccccc}
        \hline\hline
        \multicolumn{1}{l}{}&\multicolumn{1}{c}{Point}&\multicolumn{1}{c}{RBC}&\multicolumn{1}{c}{RBC}&\multicolumn{1}{c}{Sample}&\multicolumn{1}{c}{}\tabularnewline
        \multicolumn{1}{l}{}&\multicolumn{1}{c}{Estimate}&\multicolumn{1}{c}{95\% CI}&\multicolumn{1}{c}{$p$-value}&\multicolumn{1}{c}{Size}&\multicolumn{1}{c}{$h$}\tabularnewline
        \hline
\multicolumn{6}{l}{\bfseries Panel (a): Overall Average Effect} \tabularnewline
~~$\tau$&0.275&[0.176 ; 0.357]&0.000&14,622&0.151\tabularnewline
        \hline
 \multicolumn{6}{l}{\bfseries Panel (b): Heterogeneity with Discretized Income} \tabularnewline
        ~~\emph{Binary}&&&&&\tabularnewline
~~\qquad Below median&0.373&[0.284 ; 0.482]&0.000&9,454&0.188\tabularnewline
~~\qquad Above median&0.122&[-0.043 ; 0.238]&0.174&6,167&0.154\tabularnewline
        ~~\emph{By Quartile}&&&&&\tabularnewline
~~\qquad1&0.416&[0.253 ; 0.527]&0.000&4,354&0.151\tabularnewline
~~\qquad2&0.334&[0.201 ; 0.537]&0.000&4,132&0.166\tabularnewline
~~\qquad3&0.164&[-0.164 ; 0.367]&0.453&2,330&0.151\tabularnewline
~~\qquad4&0.095&[-0.070 ; 0.230]&0.297&4,495&0.204\tabularnewline
        ~~\emph{By Decile}&&&&&\tabularnewline
~~\qquad1&0.425&[0.121 ; 0.626]&0.004&1,757&0.150\tabularnewline
~~\qquad2&0.392&[0.133 ; 0.662]&0.003&1,677&0.132\tabularnewline
~~\qquad3&0.472&[0.358 ; 0.727]&0.000&1,928&0.221\tabularnewline
~~\qquad4&0.257&[0.023 ; 0.521]&0.032&1,733&0.158\tabularnewline
~~\qquad5&0.302&[0.080 ; 0.688]&0.013&1,458&0.161\tabularnewline
~~\qquad6&0.087&[-0.661 ; 0.556]&0.866&956&0.190\tabularnewline
~~\qquad7&0.288&[0.018 ; 0.535]&0.036&1,064&0.150\tabularnewline
~~\qquad8&0.067&[-0.276 ; 0.149]&0.559&912&0.150\tabularnewline
~~\qquad9&0.068&[-0.127 ; 0.526]&0.231&1,325&0.134\tabularnewline
~~\qquad10&0.124&[-0.066 ; 0.335]&0.188&2,774&0.259\tabularnewline
        \hline
 \multicolumn{6}{l}{\bfseries Panel (c): Heterogeneity with Linear Income} \tabularnewline
 
~~$\theta(0)$&0.471&[0.306 ; 0.635]&0.000&14,410&0.152\tabularnewline
~~$\xi(0)$&-0.217&[-0.385 ; -0.068]&0.005&14,410&0.152\tabularnewline
        \hline
\multicolumn{6}{l}{\bfseries Panel (d): Heterogeneity with Quadratic Income}\tabularnewline
~~$\theta(0)$&0.625&[0.231 ; 0.923]&0.001&14,410&0.152\tabularnewline
~~$\xi_1(0)$&-0.586&[-1.328 ; 0.359]&0.260&14,410&0.152\tabularnewline
~~$\xi_2(0)$&0.161&[-0.287 ; 0.513]&0.580&14,410&0.152\tabularnewline
                \hline
    \end{tabular}
    }
    \scriptsize{
    \begin{flushleft}Notes:\end{flushleft}\vspace{-0.25in}
    \begin{enumerate}[label=\normalfont(\roman*),noitemsep,leftmargin=*]
        \item Panel (a) shows the estimated RD treatment effect $\dot{\tau}$ from \eqref{eqn:reg 1}; Panel (b) shows $\widehat{\kappa}(\cdot)$ from \eqref{eqn:reg 3}--\eqref{eqn:cate hat}, where income is discretized and entered as a vector of indicator variables for each category, excluding the first category; Panel (c) shows $\widehat{\kappa}({\tt income})$ from \eqref{eqn:reg 3}--\eqref{eqn:cate hat} with income used linearly as a continuous variable, $W_i = \mathtt{income}_i$; and Panel (d) shows $\widehat{\kappa}({\tt income})$ from \eqref{eqn:reg 3}--\eqref{eqn:cate hat} with income used quadratically as a continuous variable, $W_i = (\mathtt{income}_i,\mathtt{income}_i^2)'$.
    
        \item Column ``Point Estimate'' reports local linear regression ($p=1$) with a triangular kernel MSE-optimal treatment effect point estimators, implemented using the data-driven bandwidth given in Column ``$h$''. For more details, see Theorems \ref{thm:plim} and \ref{thm:mse}, and the supplemental appendix.
    
        \item Columns ``RBC 95\% CI'' and ``RBC $p$-value'' report robust bias-corrected 95\% confidence intervals and $p$-values, respectively, implemented using the data-driven bandwidth given in Column ``$h$''. Variance Estimators are cluster-robust at the municipality level \citep[as in the original study,][column 1 of Table A.21]{Akhtari-Moreira-Trucco2022_AER}. See Theorem \ref{thm:clt-rbc}, and the supplemental appendix, for more details.
    
        \item Column ``Sample Size'' shows the effective sample size used for each estimation and robust bias-corrected inference procedure, as determined by the data-driven bandwidth reported in column ``$h$''. The total sample size is $n=26,099$.
        
        \item See the replication files at \url{https://rdpackages.github.io/replication/}, and the supplemental appendix, for more details.
    \end{enumerate}
    }
\end{table}

\clearpage

\clearpage

\bibliography{CCFPT_2025_HTERD--bib}
\bibliographystyle{jasa}


\end{document}


\title{Online Supplement for ``Treatment Effect Heterogeneity in Regression Discontinuity Designs''}
\author{Sebastian Calonico\thanks{Graduate School of Management, UC Davis.} \and
    Matias D. Cattaneo\thanks{Department of Operations Research and Financial Engineering, Princeton University.} \and
    Max H. Farrell\thanks{Department of Economics, UC Santa Barbara.} \and
    Filippo Palomba\thanks{Department of Economics, Princeton University.} \and
    Roc{\'i}o Titiunik\thanks{Department of Politics, Princeton University.}}
\maketitle

\onehalfspacing

\begin{abstract}
\noindent This supplement contains all proofs, additional results, and other technical details about estimation, inference, and identification. Section \ref{sec: introduction} describes setup and notation, states the assumptions we rely on, and introduces some auxiliary results. Section \ref{sec: main results} illustrates the main technical results. Section \ref{sec: empirical app} discusses in detail the common strategies used in empirical analyses. Section \ref{sec: proofs} contains all the proofs.
\end{abstract}
\vspace{-.1in}

\clearpage
\tableofcontents
\singlespacing
\clearpage
\section{Introduction}
\label{sec: introduction}
This section introduces the notation and links it with the one used in the main text, sets up the statistical framework, and enumerates the assumptions we rely on throughout this supplement.

\subsection{Notation}
\label{sec: notation}

In this Supplemental Appendix, we use $n$ to denote sample size and $h\equiv h_n$ to denote a bandwidth sequence where we omit the dependence on $n$ to ease notation. Moreover, $h$ denotes a generic bandwidth (e.g. $h=h_{-}$ or $h=h_{+}$ depending on the context).

\textbf{Linear algebra.} Throughout the text, $\mathbf{e}_{\nu}$ denotes a conformable vector of zeros with a 1 in its $(\nu+1)$-th element, which may take different dimensions in different places, $\mathbf{0}_k$ and $\biota_k$ are the $k$-dimensional zero and one vectors, respectively, $\bI_k$ and $\mathbf{0}_{k\times j}$ denote the $k\times k$ identity matrix and a $k\times j$ matrix of zeros, respectively, $\otimes$ indicates the Kronecker product, $\mathrm{tr}(\cdot)$ is the trace operator, and $\mathrm{diag}(\bx)$ yields a square diagonal matrix with the elements of $\bx$ on its main diagonal. With a slight abuse of notation, we denote with $\bv^k = (v_1^k,\cdots,v_n^k)^\prime$ the element-wise power for vectors $\bv\in \R^n$. The maximum and minimum of two real numbers $a$ and $b$ are denoted by $a \vee b$ and $a \wedge b$, respectively. We let $|\cdot|$ denote the Euclidean norm, $|\bA|^2 = {\sum_i\sum_j |a_{ij}|^2} = {\mathrm{tr}(\bA^\prime\bA)}$. Finally, for some $q\in\N$, with $\mathcal{C}^q$ we denote the space of functions that are $q$-times continuously differentiable.

\textbf{Asymptotic statements.} For two positive sequences $\left\{a_n\right\}_n,\left\{b_n\right\}_n$, we write $a_n = O(b_n)$ if $\exists\,M\in\R_{++}:a_n\leq M b_n$ for all large $n$, $a_n=o(b_n)$ if $\lim_{n\to\infty} a_n b_n^{-1} = 0$, and $a_n\lesssim b_n$ if there exists a constant $C$ such that $a_n\leq C b_n$ for all large $n$. For two sequences of random variables $\left\{A_n\right\}_n,\left\{B_n\right\}_n$, we write $A_n=\op(B_n)$ if $\forall\,\epsilon>0, \lim_{n\to\infty} \PP[|A_n B_n^{-1}|\geq \varepsilon]= 0$ and $A_n =\Op(B_n)$ if $\forall\,\varepsilon>0, \exists \,M,n_0\in\R_{++} : \PP[|A_nB_n^{-1}|>M]<\varepsilon,$ for $n>n_0$. We denote convergence in probability with $\cprob$ and convergence in distribution with $\cdist$. We denote (possibly multivariate) Gaussian random variable with $\mathsf{N}(\ba, \bB),$ where $\ba$ denotes the mean and $\bB$ the variance-covariance.

\textbf{Causal Model.} We now describe the population causal model. The outcome variable is
\begin{align*}
Y_i&=T_i \cdot Y_i(1)+\left(1-T_i\right) \cdot Y_i(0),
\end{align*}
with $\left(Y_i(0), Y_i(1)\right)\in\R^2$ denoting the potential outcomes and $T_i:=\indic(X_i\geq c)$ denoting treatment status. We denote the vector of covariates with $\bW_i \in \R^d, d\in\N$. We stress that, as the notation suggests, $\bW_i$ is interpreted as a pretreatment vector of covariates in the sense that $\bW_i = \bW_i(1)=\bW_i(0)$.

In sharp RD designs, $T_i=\indic\left(X_i \geq c\right)$, where $X_i\in\R$ denotes the running variable and $c\in\R$ is the cutoff. Throughout, $F_R(\cdot)$ denotes the cumulative distribution function (cdf) of a random variable $R$ and $f_R(\cdot)$ is the density of $R$ with respect to the Lebesgue measure. We denote with $f(\cdot)$ the density of $X_i$.

We further define
\begin{align*}
    \mathbf{Y}&=\left[Y_{1}, \cdots, Y_{n}\right]^{\prime},&
    \bW &= [\bW_1,\bW_2,\cdots,\bW_n]^\prime,\\
    \bmu_{W}(x)&=\E[\bW_i|X_i=x], & \bmu_{WW}(x)&=\E[\bW_i\bW_i^\prime|X_i=x], \\
    \mu_{W_\ell}(x)&=\E[W_{i\ell}|X_i=x], & \mu_{W_\ell W_j}(x)&=\E[W_{i\ell}W_{ij}|X_i=x], \quad \ell,j \in\{ 1,\cdots, d\}, \\
    \mu_{Y-}(x,\bw) &= \E[Y_i(0)\mid X_i=x, \bW_i=\bw], & \mu_{Y+}(x,\bw) &= \E[Y_i(1)\mid X_i=x, \bW_i=\bw], \\
    \mu^{(\nu)}_{Y-}(x,\bw) &= \frac{\partial^\nu}{\partial x^\nu}\E[Y_i(0)\mid X_i=x, \bW_i=\bw], & \mu_{Y+}(x,\bw) &= \frac{\partial^\nu}{\partial x^\nu}\E[Y_i(1)\mid X_i=x, \bW_i=\bw], \\
    \sigma^2_{Y-}(x,\bw) &= \V[Y_i(0)\mid X_i=x, \bW_i=\bw], & \sigma^2_{Y+}(x,\bw) &= \V[Y_i(1)\mid X_i=x, \bW_i=\bw],
    \end{align*}
and $\bsigma^2_{W}(x) =\bmu_{WW}(x)-\bmu_{W}(x)\bmu_{W}(x)^\prime$.

Let $\br_q(x)=(1,x,\ldots, x^q)^\prime$ be the polynomial basis of order $q\in\N$ and let
\[\br_{p,s}(u, \bw) = (\br_p(u)^\prime,\bw^\prime\otimes \br_s(u)^\prime)^\prime.\]
For a generic kernel function $k(\cdot)$ let
\[K(u)=\indic(u<0)k(-u) + \indic(u\geq 0)k(u)\]
and
\[K_{\bh}(u) = \indic(u<0)k_{h_{-}}(-u) + \indic(u\geq 0)k_{h_{+}}(u),\quad k_h(u) = k(u/h)/h, \quad \bh = (h_{-},h_{+})^\prime. \]
Finally, we also define the following matrices 
\[   {\bLambda}_{-,p,s} := f(c)\int\limits_{-\infty}^0 K(u) \br_p(u)\br_s(u)^\prime \de u ,\qquad {\bLambda}_{+,p,s} := f(c)\int\limits_0^\infty K(u) \br_p(u)\br_s(u)^\prime \de u,
\]
and 
\[\bXi_{-,p,s} = f(c)\int\limits_{-\infty}^0 K^2(u)\br_p\left(u\right)\br_s\left(u\right)^\prime \de u, \qquad \bXi_{+,p,s} =f(c)\int\limits_0^\infty K^2(u)\br_p\left(u\right)\br_s\left(u\right)^\prime \de u. \]

\subsection{Setup}
\label{subsec: setup}

Formally, the RD estimators with treatment interactions can be obtained from a ``long'' regression, i.e.,
\begin{align*}
\widehat{\bvartheta}_{p, s}(\bh)
&:=\argmin_{\substack{\ba_-,\ba_+ \in \mathbb{R}^{1+p}, \\ \bell_-,\bell_+ \in \mathbb{R}^{d(1+s)}}} \sum_{i=1}^n \left(Y_i-
\begin{bmatrix}\mathbf{r}_{p}\left(X_i-c\right)\\\mathbf{r}_{p}\left(X_i-c\right)\end{bmatrix}^{\prime} 
\begin{bmatrix}\ba_-\\\ba_+\end{bmatrix} -
\begin{bmatrix}(\bW_i\otimes\br_{s}(X_i-c)) \\ (\bW_i\otimes\br_{s}(X_i-c))\end{bmatrix}^{\prime} 
\begin{bmatrix}\bell_- \\ \bell_+ \end{bmatrix}\right)^2 K_{\bh}\left(X_i-c\right)  \\
\widehat{\bvartheta}_{ p, s}(\bh)
&=
\begin{bmatrix}
\widehat{\bvartheta}_{-, p, s}(h_-)\\
\widehat{\bvartheta}_{+, p, s}(h_+)
\end{bmatrix}
=
\begin{bmatrix}
\widehat{\balpha}_{-, p}(h_-) \\
\widehat{\blambda}_{-, s}(h_-) \\
\widehat{\balpha}_{+, p}(h_+) \\
\widehat{\blambda}_{+, s}(h_+)
\end{bmatrix}, \qquad p,s\in\N,
\end{align*}
where
\begin{align*}
\widehat{\balpha}_{-,p}(h) =(\widehat{\alpha}_{-,0}, \cdots, \widehat{\alpha}_{-,p})',\qquad
\widehat{\balpha}_{+,p}(h) =(\widehat{\alpha}_{+,0}, \cdots, \widehat{\alpha}_{+,p})',
\end{align*}
and
\begin{align*}
\widehat{\blambda}_{-,s}(h) &= (\widehat{\lambda}_{-,10}(h),\cdots, \widehat{\lambda}_{-,1s}(h), \cdots, \widehat{\lambda}_{-,d0}(h),\cdots, \widehat{\lambda}_{-,ds}(h))', \\    
\widehat{\blambda}_{+,s}(h) &= (\widehat{\lambda}_{+,10}(h),\cdots, \widehat{\lambda}_{+,1s}(h), \cdots, \widehat{\lambda}_{+,d0}(h),\cdots, \widehat{\lambda}_{+,ds}(h))'.    
\end{align*}

Note that the vector $\widehat{\blambda}_{+,s}(h)$ (similarly, $\widehat{\blambda}_{-,s}(h)$) stacks first all the coefficients of the polynomial basis of order $s$ for the first covariate, then for the second, and so on until the $d$-th covariate.

In what follows, we define $m:=d(1+s)$. The regression above has orthogonal regressors. Indeed, it has a set of regressors that pertains to estimation to the left of the cutoff, i.e., $(\indic (X_i<c)\br_{p}(X_i-c)^\prime, \indic (X_i\leq c)\bW_i^\prime\otimes\br_{s}(X_i-c)^\prime)^\prime$, and a second set of regressors that refers to estimation to the right of the cutoff, i.e., $(\indic (X_i\geq c)\br_{p}(X_i-c)^\prime, \indic (X_i\geq c)\bW_i^\prime\otimes\br_{s}(X_i-c)^\prime)^\prime$. Accordingly, by the partitioned regression theorem, the least squares coefficients in the ``long'' regression are equivalent to the least squares coefficients in the following ``short'' regressions:
\begin{align*}
\widehat{\bvartheta}_{-, p,s}(h)=\left[\begin{array}{c}
\widehat{\balpha}_{-, p}(h) \\
\widehat{\blambda}_{-, s}(h)
\end{array}\right]=\argmin_{\substack{\mathbf{a} \in \mathbb{R}^{1+p},\\ \bell \in \mathbb{R}^m}} \sum_{i=1}^n \indic\left(X_i<c\right)\left(Y_i-\mathbf{r}_p\left(X_i-c\right)^{\prime} \mathbf{a}-\left(\bW_i\otimes\br_s(X_i-c)\right)^{\prime} \bell\right)^2 K_h\left(X_i-c\right), \\
\widehat{\bvartheta}_{+, p, s}(h)=\left[\begin{array}{c}
\widehat{\balpha}_{+, p}(h) \\
\widehat{\blambda}_{+, s}(h)
\end{array}\right]=\argmin_{\substack{\mathbf{a} \in \mathbb{R}^{1+p},\\ \bell \in \mathbb{R}^m}}  \sum_{i=1}^n \indic\left(X_i \geq c\right)\left(Y_i-\mathbf{r}_p\left(X_i-c\right)^{\prime} \mathbf{a}-(\bW_i\otimes\br_s(X_i-c))^{\prime} \bell\right)^2 K_h\left(X_i-c\right),
\end{align*}
which can be equivalently written in matrix form as
\begin{align*}
    \widehat{\bvartheta}_{-, p,s}(h)=\bH_{p,s}^{-1}(h)\widehat{\bGamma}^{-1}_{-,p,s}(h)\widehat{\bUpsilon}_{-,p,s}(h),\qquad
    \widehat{\bvartheta}_{+, p,s}(h)=\bH_{p,s}^{-1}(h)\widehat{\bGamma}^{-1}_{+,p,s}(h)\widehat{\bUpsilon}_{+,p,s}(h),
\end{align*}
where the Hessian and score matrices are defined as
\begin{align*}
    \widehat{\bGamma}_{-,p,s}(h)=\bR_{p,s}(h)^\prime\bK_-(h)\bR_{p,s}(h)/n, \qquad &\widehat{\bUpsilon}_{-,p,s}(h) =\bR_{p,s}(h)^\prime\bK_-(h)\bY/n, \\
    \widehat{\bGamma}_{+,p,s}(h)=\bR_{p,s}(h)^\prime\bK_+(h)\bR_{p,s}(h)/n, \qquad &\widehat{\bUpsilon}_{+,p,s}(h) =\bR_{p,s}(h)^\prime\bK_+(h)\bY/n, \\
\end{align*}
where 
\[\bR_{p}(h)=\begin{bmatrix}
    \br_p\left(\frac{X_1-c}{h}\right)^\prime \\
    \br_p\left(\frac{X_2-c}{h}\right)^\prime \\
    \vdots \\
    \br_p\left(\frac{X_n-c}{h}\right)^\prime \\
\end{bmatrix}_{n \times (1+p)},\quad \bR_{p,s}(h)=\begin{bmatrix}
    \br_p\left(\frac{X_1-c}{h}\right)^\prime & \bW_1'\otimes\br_s\left(\frac{X_1-c}{h}\right)^\prime  \\
    \br_p\left(\frac{X_2-c}{h}\right)^\prime & \bW_2'\otimes\br_s\left(\frac{X_2-c}{h}\right)^\prime \\
    \vdots & \vdots \\
    \br_p\left(\frac{X_n-c}{h}\right)^\prime & \bW_n'\otimes\br_s\left(\frac{X_n-c}{h}\right)^\prime \\
\end{bmatrix}_{n \times (1+p + m)},\]
with the scaling and kernel matrices defined as
\[\bH_{p,s}(h) = 
\begin{bmatrix}
\bH_p(h) & \mathbf{0}_{(1+p)\times m} \\
\mathbf{0}_{m\times (1+p)} & \bI_d\otimes \bH_s(h) \\
\end{bmatrix}_{(1+p + m)\times(1+p + m)}, \qquad \bH_q(h) = \text{diag}(h^\ell:\ell=0,\ldots,q),\]
\begin{align*}
    \bK_-(h) &= \text{diag}\left(\indic ( X_i< c)k_h(X_i-c): i = 1, \ldots, n\right), \quad
\bK_+(h) = \text{diag}\left(\indic ( X_i\geq c)k_h(X_i-c): i = 1, \ldots, n\right).
\end{align*}

In what follows, we mostly refer to the RD estimator to the right of the cutoff, $\widehat{\bvartheta}_{+, p, s}(h)$, but everything follows symmetrically for the other RD estimator to the left of the cutoff, $\widehat{\bvartheta}_{-, p, s}(h)$. 

We focus our attention on estimators of the following form:
\[\widehat{\chi}_{\nu,p,s}(\bh):=\cbe_{\nu}^\prime(\widehat{\bvartheta}_{+,p,s}(h_+)-\widehat{\bvartheta}_{-,p,s}(h_-)),\qquad \nu\in\{0,1,\ldots,p\wedge s\},
\]
where for fixed $\bw\in\R^d$
\[\cbe_\nu \equiv \cbe_{\nu}(\bw) := \begin{bmatrix}
    \nu\bI_{p+1}\be_{\nu} \\ \nu!\bw \otimes \bI_{s+1}\be_{{\nu}}
\end{bmatrix}\]
is the vector which extracts the $(\nu+1)$-th term in the polynomial basis in $(X_i-c)$ and the $(\nu+1)$-th terms in the interaction between such basis and $\bW_i$. The identity matrices are there just to stress the different dimensions $\be_\nu$ takes in different places. Most of the time, we suppress the dependence of $\cbe_\nu$ on $\bw$ to avoid making the notation even more cumbersome and because in the case of orthogonal covariates, the only sensible evaluation point would be $\bw = \biota_d$.\footnote{This case corresponds to situations where each component is binary, $\bW_i \in \{0,1\}^d$, and at most one component takes the value one, implying that $\bW_i'\bW_i \leq 1$.} The evaluation point $\bw$ would only turn out to be useful when some elements in $\bW_i$ are continuous variables, as the following example clarifies.
\begin{example}
Suppose $d=1, W_i \in\{0,1\}$ and $\nu=0$, then
\[\cbe_{0}^\prime\widehat{\bvartheta}_{+,p,s}(h) = \be_{0}^\prime\widehat{\balpha}_{+,p}(h) + \be_{0}^\prime\widehat{\blambda}_{+,s}(h)= \underbrace{\widehat{\alpha}_{+,0}(h)}_{\text{coefficient on } \indic(X_i\geq c)} + \underbrace{\widehat{\lambda}_{+,10}(h)}_{\text{coefficient on } W_i\cdot \indic(X_i\geq c) },\]
and
\[\cbe_{0}^\prime\widehat{\bvartheta}_{-,p,s}(h) = \be_{0}^\prime\widehat{\balpha}_{-,p}(h) + \be_{0}^\prime\widehat{\blambda}_{-,s}(h)= \underbrace{\widehat{\alpha}_{-,0}(h)}_{\text{coefficient on } \indic(X_i< c)} + \underbrace{\widehat{\lambda}_{-,10}(h)}_{\text{coefficient on } W_i\cdot \indic(X_i< c) }.\]

If instead $\nu=1$, then 
\[\cbe_{1}^\prime\widehat{\bvartheta}_{+,p,s}(h) = \be_{1}^\prime\widehat{\balpha}_{+,p}(h) + \be_{1}^\prime\widehat{\blambda}_{+,s}(h)= \underbrace{\widehat{\alpha}_{+,1}(h)}_{\text{coefficient on } \indic(X_i\geq c)\cdot(X_i-c)} + \underbrace{\widehat{\lambda}_{+,11}(h)}_{\text{coefficient on } W_i\cdot \indic(X_i\geq c)\cdot(X_i-c) },\]
and
\[\cbe_{1}^\prime\widehat{\bvartheta}_{-,p,s}(h) = \be_{1}^\prime\widehat{\balpha}_{-,p}(h) + \be_{1}^\prime\widehat{\blambda}_{-,s}(h)= \underbrace{\widehat{\alpha}_{-,1}(h)}_{\text{coefficient on } \indic(X_i< c)\cdot(X_i-c)} + \underbrace{\widehat{\lambda}_{-,11}(h)}_{\text{coefficient on } W_i\cdot \indic(X_i< c)\cdot(X_i-c) }.\]

Now, suppose that $W_i$ is a continuous covariate and $\nu = 0$. In this case, the evaluation point $w=w_0$ does matter, and so
\[\cbe_{0}^\prime\widehat{\bvartheta}_{+,p,s}(h) = \be_{0}^\prime\widehat{\balpha}_{+,p}(h) + w_0\be_{0}^\prime\widehat{\blambda}_{+,s}(h)= \underbrace{\widehat{\alpha}_{+,0}(h)}_{\text{coefficient on } \indic(X_i\geq c)} + \underbrace{\widehat{\lambda}_{+,10}(h)}_{\substack{\text{coefficient on } \\ W_i\cdot \indic(X_i\geq c)} }\cdot\, w_0,\]
and
\[\cbe_{0}^\prime\widehat{\bvartheta}_{-,p,s}(h) = \be_{0}^\prime\widehat{\balpha}_{-,p}(h) + \be_{0}^\prime\widehat{\blambda}_{-,s}(h)= \underbrace{\widehat{\alpha}_{-,0}(h)}_{\text{coefficient on } \indic(X_i< c)} + \underbrace{\widehat{\lambda}_{-,10}(h)}_{\substack{\text{coefficient on } \\ W_i\cdot \indic(X_i< c)} }\cdot\, w_0.\]
\end{example}

Moreover, let
\begin{align*}
    \bvartheta_{-,p,s}^\star(h)&:= \argmin_{\bt\in\R^{1+p+m}}\E\left[\indic(X_i< c)K\left(\frac{X_i-c}{h}\right)(Y_i - \br_{p,s}(X_i-c,\bW_i)^\prime\bt)^2\right],\\
    \bvartheta_{+,p,s}^\star(h)&:= \argmin_{\bt\in\R^{1+p+m}}\E\left[\indic(X_i\geq c)K\left(\frac{X_i-c}{h}\right)(Y_i - \br_{p,s}(X_i-c,\bW_i)^\prime\bt)^2\right],
\end{align*}
be the fixed-$h$ best linear mean square error predictors of $\E[Y_i(d)\mid X_i=c, \bW_i], d\in\{0,1\}$, i.e.,
\begin{align*}
    \bvartheta^\star_{-,p,s}(h) &= \bH_{p,s}^{-1}(h)\E\left[\indic(X_i< c) K\left(\frac{X_i-c}{h}\right)\br_{p,s}\left(\frac{X_i-c}{h},\bW_i\right)\br_{p,s}\left(\frac{X_i-c}{h},\bW_i\right)^\prime\right]^{-1} \times \\
    &\qquad\qquad\qquad\qquad\qquad\qquad \E\left[\indic(X_i< c) K\left(\frac{X_i-c}{h}\right)\br_{p,s}\left(\frac{X_i-c}{h},\bW_i\right)\mu_{Y-}(X_i,\bW_i)\right] \\
    &=\bH_{p,s}^{-1}(h)\E[\widehat{\bGamma}_{-,p,s}(h)]^{-1}\E[\widehat{\bUpsilon}_{-,p,s}(h)],
\end{align*}
and
\begin{align*}
    \bvartheta^\star_{+,p,s}(h) &= \bH_{p,s}^{-1}(h)\E\left[\indic(X_i\geq c) K\left(\frac{X_i-c}{h}\right)\br_{p,s}\left(\frac{X_i-c}{h},\bW_i\right)\br_{p,s}\left(\frac{X_i-c}{h},\bW_i\right)^\prime\right]^{-1} \times \\
    &\qquad\qquad\qquad\qquad\qquad\qquad \E\left[\indic(X_i\geq c) K\left(\frac{X_i-c}{h}\right)\br_{p,s}\left(\frac{X_i-c}{h},\bW_i\right)\mu_{Y+}(X_i,\bW_i)\right] \\
    &=\bH_{p,s}^{-1}(h)\E[\widehat{\bGamma}_{+,p,s}(h)]^{-1}\E[\widehat{\bUpsilon}_{+,p,s}(h)].
\end{align*}

\subsection{Assumptions}

In what follows, we state all the assumptions we rely on to prove the results in this supplemental appendix. Assumption 1 in the main text contains Assumptions \ref{ass: sampling}, \ref{ass: density running variable}, \ref{ass: DGP}, and \ref{ass: kernel}, whereas Assumption 2 coincides with \ref{ass: conditional expectation}.

\begin{assumption}[Sampling]\label{ass: sampling}
    $\{(Y_i,X_i,\bW_i^\prime)^\prime\}_{i=1}^n$ is a random sample.
\end{assumption}

\begin{assumption}[Density of Running Variable] \label{ass: density running variable}
The Lebesgue density of $X_i$ is continuous, bounded, and bounded away from zero.
\end{assumption}

\begin{assumption}[DGP]\label{ass: DGP}
For $ t\in \{0,1\}$ and $\mathsf{k}\in \R_{++}, $ with $-\mathsf{k}<c<\mathsf{k}$ and for all $x\in [-\mathsf{k},\mathsf{k}], \bw\in\R^d:$
\begin{enumerate}[label=(\alph*), ref={SA3\alph*}]
    \item $\E[\bW_i\mid X_i=x]$ is continuous and $\E[\bW_i\bW_i'\mid X_i=x]$ is continuous and invertible. \label{ass: SRD continuous W}
    \item $\E[Y_i(t)\mid X_i=x, \bW_i=\bw]$is $q$-times continuously differentiable in $x$ and continuous in $\bw$ for $q\geq 1$. \label{ass: SRD continuous Y}
    \item $\V[Y_i(t)|X_i=x]$ and $\V[Y_i(t)|X_i=x, \bW_i=\bw]$ are continuous and bounded away from zero. \label{ass: SRD continuous variance}
    \item $\E[|Y_i(t)|^4|X_i=x, \bW_i=\bw]$ is continuous in both arguments. \label{ass: continuous fourth moment of Y}
    \item $\E[|\bW_i|^4|X_i=x]$ is continuous.  \label{ass: SRD continuous fourth moment of W}
    \item $\E[\bW_i \V[Y_i(t)|X_i,\bW_i]\mid X_i=x]$ and $\E[\bW_i\bW_i^\prime\V[Y_i(t)|X_i,\bW_i]\mid X_i=x]$ are continuous.  \label{ass: SRD continuous covariances}
    \item ${\bGamma}_{-,p,s}$ and ${\bGamma}_{+,p,s}$ in Equation \eqref{eq: plim hessian} are positive definite matrices. \label{ass: pd gram matrix}
\end{enumerate}  
\end{assumption}

\begin{assumption}[Identification]
\label{ass: conditional expectation} For $\mathsf{k}\in \R, $ with $-\mathsf{k}<c<\mathsf{k}$ and for all $x\in [-\mathsf{k},\mathsf{k}], \bw\in\R^d$ the expectation of $Y_i(t), t\in\{0,1\}$ conditional on $(X_i,\bW_i)$ is
\begin{align*}
    \mu_-(x, \bw) = \alpha_{-}(x) + \blambda_{-}(x)^\prime \bw, \qquad
    \mu_+(x, \bw) = \alpha_{+}(x) + \blambda_{+}(x)^\prime \bw,
\end{align*}
where $\blambda_{-}(x):=(\lambda_{-,1}(x), \cdots, \lambda_{-,d}(x))^\prime, \blambda_{+}(x):=(\lambda_{+,1}(x), \cdots, \lambda_{+,d}(x))^\prime,$ and $\alpha_{-}(x),\alpha_{+}(x)\in\mathcal{C}^{q},$ and  $\lambda_{-,\ell}(x),\lambda_{+,\ell}(x),\in\mathcal{C}^{q}$ for all $\ell\in\{1,\cdots,d\}$ for some $q\in\N$.    
\end{assumption}

\begin{assumption}[Kernel] \label{ass: kernel}
The kernel function $k(\cdot):[0,1]\to \mathbb{R}$ is continuous and nonnegative. 
\end{assumption}

In words, under Assumption \ref{ass: sampling} the sample is composed of independent draws from an underlying population of interest. Assumptions \ref{ass: density running variable}, \ref{ass: SRD continuous W}, \ref{ass: SRD continuous variance}, and \ref{ass: SRD continuous covariances} are technical conditions on the data-generating process (DGP) we rely on when we establish convergence rates and characterize the probability limits of interest. Assumption \ref{ass: pd gram matrix} is a standard assumption in least squares which requires the probability limit of the Gram matrix to be positive definite. Assumption \ref{ass: SRD continuous W} also establishes that $\bW$ can be thought of as a vector of pretreatment covariates. Assumption \ref{ass: SRD continuous Y} is necessary to Taylor expand the conditional expectation of the potential outcomes and characterize the smoothing bias. Assumptions \ref{ass: continuous fourth moment of Y} and \ref{ass: SRD continuous fourth moment of W} are standard bounded absolute higher-order moment conditions that make the Lyapunov condition satisfied and allow us to invoke a Lindeberg-Feller central limit theorem to show asymptotic normality of the RD estimator. Assumption \ref{ass: conditional expectation} imposes a partially linear (in $\bW$) structure --locally at the cutoff-- on the population conditional expectation of the potential outcomes which we rely on when attaching a causal interpretation to the probability limit of the RD estimator. Assumption \ref{ass: kernel} states standard technical conditions on the kernel used in the local polynomial regression.

\subsection{Mapping between Main Text and Supplement}

In the main text, we present all the results using the long regression with $p=s=1$
\begin{align}\label{eq: long regression interacted}
    Y_i\quad \text{onto}\quad (\br_p(X_i-c)', T_i\br_p(X_i-c)', \bW_i'\otimes\br_s(X_i-c)', T_i\bW_i'\otimes\br_s(X_i-c)')',
\end{align}
which aligns with the classical way of thinking about linear regressions. Indeed, this representation allows us to readily interpret the coefficients on the terms interacted with $T_i$ as specific differences between treated and control groups. 

In this supplement, we set up the problem as the long regression 
\begin{align}\label{eq: long regression interacted 2}
Y_i\quad \text{onto}\quad((1-T_i)\br_p(X_i-c)', T_i\br_p(X_i-c)', (1-T_i)\bW_i'\otimes\br_s(X_i-c)', T_i\bW_i'\otimes\br_s(X_i-c)')'.
\end{align}
This is equivalent to \eqref{eq: long regression interacted} in that the predicted values and the residuals of these two regressions are the same. However, the regression coefficients in \eqref{eq: long regression interacted 2} have a different interpretation than those in \eqref{eq: long regression interacted}. For example, the coefficients on the terms interacted with $T_i$ do not describe differences between treated and control groups but rather capture only moments of the treated population. On the one hand, \eqref{eq: long regression interacted} allows coefficients to be directly interpreted as long as differences between treated and control are of interest. On the other hand, as we already stressed in Section \ref{subsec: setup}, \eqref{eq: long regression interacted 2} can be written as two separate ``short" regressions with orthogonal design matrices, making statements and proofs less cumbersome from a notational standpoint. For this reason, we rely on \eqref{eq: long regression interacted} in the main text and on \eqref{eq: long regression interacted 2} in this supplement.

Table \ref{tab: mapping notation} maps the notation used in this supplement with the one used in the main paper. Dependence on $h$ is omitted to simplify notation. For the MSE constants, we will use the relationship
\[\widehat{\boldsymbol{\varsigma}}(\bh) = \bM\widehat{\bvartheta}_{1,1}(\bh) =\bM
\begin{bmatrix}
\widehat{\bvartheta}_{-, p, s}(h_-)\\
\widehat{\bvartheta}_{+, p, s}(h_+)
\end{bmatrix}, \qquad 
\bM := 
\begin{bmatrix}
-\be_0'\bI_{p+1}   & \mathbf{0}'_{m} \\
\mathbf{0}_{d\times (p+1)} & -(\biota_d\otimes\bI_{s+1}\be_0)' \\
\be_0'\bI_{p+1}   & \mathbf{0}'_{m} \\
\mathbf{0}_{d\times (p+1)} & (\biota_d\otimes\bI_{s+1}\be_0)'
\end{bmatrix}.\]

\begin{table}[!ht]
    \centering
    \caption{Mapping between main text and supplement notation.}
    \begin{tabular}{ccc}
        \toprule\toprule
       & Main  & Supplement \\
       \midrule
       \textit{Estimands} & & \\
        & $\alpha(\cdot)$ & $\alpha_-(\cdot)$ \\
        & $\theta(\cdot)$ & $\alpha_+(\cdot)-\alpha_-(\cdot)$ \\
        & $\blambda(\cdot)$ & $\blambda_-(\cdot)$ \\
        & $\bxi(\cdot)$ & $\blambda_+(\cdot)-\blambda_-(\cdot)$ \\
        & $\boldsymbol{\varsigma}$ & $(\alpha_+(0)-\alpha_-(0),\blambda_+(0)'-\blambda_-(0)')'$\\
        \textit{Estimators} & & \\
        & $\widehat{\alpha}$ & $\be_0'\widehat{\balpha}_-$ \\
        & $\widehat{\theta}$ & $\be_0'(\widehat{\balpha}_+-\widehat{\balpha}_-)$ \\
        & $\widehat{\blambda}$ & $(\biota_d\otimes\bI_{s+1}\be_0)'\widehat{\blambda}_-$\\
        & $\widehat{\bxi}$ & $(\biota_d\otimes\bI_{s+1}\be_0)'(\widehat{\blambda}_+-\widehat{\blambda}_-)$\\
        & $\widehat{\boldsymbol{\varsigma}}$ & $\cbe_0'(\widehat{\bvartheta}_{+,1,1}-\widehat{\bvartheta}_{-,1,1})$\\
        & $\widehat{\omega}_1$ & $\be_1'\widehat{\balpha}_-$ \\
        & $\widehat{\omega}_2$ & $\be_1'(\widehat{\balpha}_+-\widehat{\balpha}_-)$ \\
        & $\widehat{\bomega}_3$ & $(\biota_d\otimes\bI_{s+1}\be_1)'\widehat{\blambda}_-$ \\
        & $\widehat{\bomega}_4$ & $(\biota_d\otimes\bI_{s+1}\be_1)'(\widehat{\blambda}_+-\widehat{\blambda}_-)$ \\
        \textit{Constants} & & \\
        & $\mathsf{B}_{\boldsymbol{\mathsf{s}}}$ & $\mathsf{s}'\bM(\bB_{+,1,1}^{[0]} + \bB_{+,1,1}^{[1]}-\bB_{-,1,1}^{[0]} + \bB_{-,1,1}^{[1]})$ \\
        & $\mathsf{V}_{\boldsymbol{\mathsf{s}}}$ & $\mathsf{s}'\bM({\bGamma}_{+,1,1}^{-1} \mathbf{V}_{+,1,1}
    {{\bGamma}_{+,1,1}^{-1\prime}} + {\bGamma}_{-,1,1}^{-1} \mathbf{V}_{-,1,1}
    {{\bGamma}_{-,1,1}^{-1\prime}})\bM'\mathsf{s}$ \\
        \bottomrule\bottomrule
    \end{tabular}
    \label{tab: mapping notation}
\end{table}

We conclude this section by mapping the results in the main text with the proofs in the supplement:
\begin{itemize}
    \item Theorem 1 is a particular case of the results proved in Section \ref{subsec: consistency} with $p=s=1$ and $\nu=0$;
    \item Theorem 2 follows from the results in Section \ref{subsec: mse expansions} with $p=s=1, \nu=0,$ and $\cbe_\nu=\bs$;
    \item Theorem 3 is a particular case of Corollary \ref{cor: asympt distr under sa5} with $p=s=1$ and $\nu=0$.
\end{itemize}

\subsection{Auxiliary Lemmas and Quantities}

In this subsection, we introduce a series of auxiliary lemmas and quantities we rely upon in the rest of the supplemental appendix. We heavily rely on the next lemma for asymptotic statements.
\begin{lemma}\label{lemma: markov}
Let $\bA_n$ be a sequence of random matrices with finite first two moments. Then
\[\bA_n = \E[\bA_n] + \Op(|\V[\bA_n]|^{1/2}).\]
\end{lemma}
\begin{flushright}
[\hyperref[proof: markov]{Proof}]
\end{flushright}

The next lemma handles the Gram matrices 
\[\widehat{\bLambda}_{-,p,s}(h) = \bR_p(h)^\prime\bK_-(h)\bR_s(h)/n, \qquad \widehat{\bLambda}_{+,p,s}(h) = \bR_p(h)^\prime\bK_+(h)\bR_s(h)/n,\]
and shows the object concentrates around in probability and the rate at which such concentration occurs.
\begin{lemma}\label{lemma: hessian}
Let Assumptions \ref{ass: sampling}, \ref{ass: density running variable}, and \ref{ass: kernel} hold with $\mathsf{k} \geq h$.
If $nh\to\infty$ and $h\to 0,$ then
$$
\widehat{\bLambda}_{-, p, s}(h)=\widetilde{\bLambda}_{-, p, s}(h)+\Op(1/\sqrt{nh}), \qquad  \widehat{\bLambda}_{+, p, s}(h)=\widetilde{\bLambda}_{+, p, s}(h)+\Op(1/\sqrt{nh}),
$$
with
\begin{align*}
\widetilde{\bLambda}_{-, p, s}(h)&:= \mathbb{E}[\widehat{\bLambda}_{-,p, s}(h)] = \int\limits_{-\infty}^0K(u)\br_{p}(u)\br_s(u)^\prime f(uh+c)\de u, \\ 
\widetilde{\bLambda}_{+, p, s}(h) &:= \mathbb{E}[\widehat{\bLambda}_{+, p, s}(h)] =\int\limits_0^\infty K(u)\br_{p}(u)\br_s(u)^\prime f(uh+c)\de u.
\end{align*}
\end{lemma}
\begin{flushright}
[\hyperref[proof: hessian]{Proof}]
\end{flushright}

In the next lemma, we give conditions for the asymptotic invertibility of $\widehat{\bGamma}_{-,p,s}(h)$ and $\widehat{\bGamma}_{+,p,s}(h)$, thereby making local polynomial estimators well-defined in large samples.

\begin{lemma}
    \label{lemma: grand hessian}
    Let Assumptions \ref{ass: sampling}, \ref{ass: density running variable}, \ref{ass: DGP}, and \ref{ass: kernel} hold with $\mathsf{k} \geq h$. 
    If $nh\to\infty$ and $h\to0$, then
    \begin{align*}
        \widehat{\bGamma}_{-,p,s}(h) &= \widetilde{\bGamma}_{-,p,s}(h) + \Op(1/\sqrt{nh}),\qquad 
        \widehat{\bGamma}_{+,p,s}(h) = \widetilde{\bGamma}_{+,p,s}(h) + \Op(1/\sqrt{nh}),
    \end{align*}
    where
    \begin{align*}
        \widetilde{\bGamma}_{-,p,s}(h):= 
    \begin{bmatrix}
        \widetilde{\bLambda}_{-,p,p}(h) & \widetilde{\bG}_{-,2}(h) \\
        \widetilde{\bG}_{-,2}(h)^\prime &   \widetilde{\bG}_{-, 3}(h)
    \end{bmatrix}, \qquad 
    \widetilde{\bGamma}_{+,p,s}(h):=\begin{bmatrix}
        \widetilde{\bLambda}_{+,p,p}(h) & \widetilde{\bG}_{+,2}(h) \\
        \widetilde{\bG}_{+,2}(h)^\prime &   \widetilde{\bG}_{+, 3}(h)
    \end{bmatrix},
    \end{align*}
    where
    \begin{align*}
         \widetilde{\bG}_{-,2}(h)&=\int\limits_{-\infty}^0 K\left(u\right)\left[ \bmu_{W}(uh+c)^\prime\otimes \br_p\left(u\right)\br_s\left(u\right)^\prime\right]f(uh+c)\de u,\\
         \widetilde{\bG}_{+,2}(h)&=\int\limits_{0}^\infty K(u)\left[ \bmu_{W}(uh+c)^\prime\otimes \br_p\left(u\right)\br_s\left(u\right)^\prime\right]f(uh+c)\de u,\\
         \widetilde{\bG}_{-,3}(h)&=\int\limits_{-\infty}^0 K\left(u\right)\left[ \bmu_{WW}(uh+c)\otimes \br_s\left(u\right)\br_s\left(u\right)^\prime\right]f(uh+c)\de u,\\
         \widetilde{\bG}_{+,3}(h)&=\int\limits_{0}^\infty K\left(u\right)  \left[\bmu_{WW}(uh+c)\otimes \br_s\left(u\right)\br_s\left(u\right)^\prime\right]f(uh+c)\de u.
    \end{align*}
Last,
\[\widetilde{\bGamma}_{-,p,s}(h) = \bGamma_{-,p,s}\{1+o(1)\}, \qquad \widetilde{\bGamma}_{+,p,s}(h) = \bGamma_{+,p,s}\{1+o(1)\}, \]
where
\begin{align}
\label{eq: plim hessian}
\bGamma_{-,p,s} = \begin{bmatrix}
    \bLambda_{-,p,p} & \bG_{-,2} \\
    \bG_{-,2}' & \bG_{-,3}
\end{bmatrix},\qquad \bGamma_{+,p,s} = \begin{bmatrix}
    \bLambda_{+,p,p} & \bG_{+,2} \\
    \bG_{+,2}' & \bG_{+,3}
\end{bmatrix},
\end{align}
with
\begin{alignat*}{3}
    {\bLambda}_{-,p,s} &= f(c)\int\limits_{-\infty}^0 K(u) \br_p(u)\br_s(u)^\prime \de u ,\qquad &{\bLambda}_{+,p,s} &= f(c)\int\limits_0^\infty K(u) \br_p(u)\br_s(u)^\prime \de u \\
    {\bG}_{-,2} &= \bmu_{W}^\prime \otimes {\bLambda}_{-,p,s},\qquad &{\bG}_{+,2} &= \bmu_{W}^\prime \otimes {\bLambda}_{+,p,s}, \\
    {\bG}_{-,3} &= \bmu_{WW} \otimes {\bLambda}_{-,s,s},\qquad &{\bG}_{+,3} &= \bmu_{WW} \otimes {\bLambda}_{+,s,s}.
\end{alignat*}
\end{lemma}
\begin{flushright}
[\hyperref[proof: grand hessian]{Proof}]
\end{flushright}

To have a more compact notation, we introduce the following quantities for integers $p,s,a\in \N_0$:
\begin{align*}
    \widehat{\bzeta}_{-,p,s,a}(h) &= \frac{1}{nh}\sum_{i=1}^n\indic(X_i< c)K\left(\frac{X_i-c}{h}\right)\br_{p,s}\left(\frac{X_i-c}{h},\bW_i\right)\left(\frac{X_i-c}{h}\right)^{a+1}, \\
    \widehat{\bzeta}_{+,p,s,a}(h) &= \frac{1}{nh}\sum_{i=1}^n\indic(X_i\geq c)K\left(\frac{X_i-c}{h}\right)\br_{p,s}\left(\frac{X_i-c}{h},\bW_i\right)\left(\frac{X_i-c}{h}\right)^{a+1}, \\
    \widehat{\bvarphi}_{-,p,s,a}(h) &= \frac{1}{nh}\sum_{i=1}^n\indic(X_i< c)K\left(\frac{X_i-c}{h}\right)\br_{p,s}\left(\frac{X_i-c}{h},\bW_i\right)\bW_i^\prime\left(\frac{X_i-c}{h}\right)^{a+1},\\
    \widehat{\bvarphi}_{+,p,s,a}(h) &= \frac{1}{nh}\sum_{i=1}^n\indic(X_i\geq c)K\left(\frac{X_i-c}{h}\right)\br_{p,s}\left(\frac{X_i-c}{h},\bW_i\right)\bW_i^\prime\left(\frac{X_i-c}{h}\right)^{a+1}.
    \end{align*}

The following two lemmas characterize the asymptotic properties of $\widehat{\bzeta}_{-,p,s,a}(h), \widehat{\bzeta}_{+,p,s,a}(h), \widehat{\bvarphi}_{-,p,s,a}(h)$, and $\widehat{\bvarphi}_{+,p,s,a}(h)$. These results are used below to guarantee that the ``constant terms'' in the MSE expansions are asymptotically well-defined.

\begin{lemma}\label{lemma: vartheta hat order}
    Let Assumptions \ref{ass: sampling}, \ref{ass: density running variable}, \ref{ass: DGP}, and \ref{ass: kernel} hold with $\mathsf{k} \geq h$. If $nh\to\infty$ and $h\to 0 $, then
\begin{alignat*}{3}
    \widehat{\bzeta}_{-,p,s,a}(h) &= \widetilde{\bzeta}_{-,p,s,a}(h) + \Op(1/\sqrt{nh}), \qquad &\widehat{\bzeta}_{+,p,s,a}(h) &= \widetilde{\bzeta}_{+,p,s,a}(h) + \Op(1/\sqrt{nh}),\\
    \widehat{\bvarphi}_{-,p,s,a}(h) &= \widetilde{\bvarphi}_{-,p,s,a}(h) + \Op(1/\sqrt{nh}), \qquad &\widehat{\bvarphi}_{+,p,s,a}(h) &= \widetilde{\bvarphi}_{+,p,s,a}(h) + \Op(1/\sqrt{nh}),    
\end{alignat*}
with 
\begin{align*}
    \widetilde{\bzeta}_{-,p,s,a}(h)&:=\E[\widehat{\bzeta}_{-,p,s,a}(h)] =\int\limits_{-\infty}^0 K(u)\br_{p,s}(u,\bmu_{W}(uh+c))u^{a+1}f(uh+c)\de u, \\
    \widetilde{\bzeta}_{+,p,s,a}(h)&:=\E[\widehat{\bzeta}_{+,p,s,a}(h)]=\int\limits_0^\infty K(u)\br_{p,s}(u,\bmu_{W}(uh+c))u^{a+1} f(uh+c)\de u,\\
    \widetilde{\bvarphi}_{-,p,s,a}(h)&:= E[\widehat{\bvarphi}_{-,p,s,a}(h)]=\int\limits_{-\infty}^0 K(u) \begin{bmatrix}
    \br_p(u)\bmu_{W}(uh+c)^\prime \\
    \bmu_{WW}(uh+c)\otimes \br_s(u)
\end{bmatrix} u^{a+1} f(uh+c)\de u, \\
    \widetilde{\bvarphi}_{+,p,s,a}(h)&:= \E[\widehat{\bvarphi}_{+,p,s,a}(h)]=\int\limits_0^\infty K(u) \begin{bmatrix}
    \br_p(u)\bmu_{W}(uh+c)^\prime \\
    \bmu_{WW}(uh+c)\otimes \br_s(u)
\end{bmatrix} u^{a+1} f(uh+c)\de u.
\end{align*}

Last
\begin{alignat*}{3}
    \widetilde{\bzeta}_{-,p,s,a}(h)&={\bzeta}_{-,p,s,a}\{1+o(1)\}, \qquad &\widetilde{\bzeta}_{+,p,s,a}(h)&={\bzeta}_{+,p,s,a}\{1+o(1)\}, \\
    \widetilde{\bvarphi}_{-,p,s,a}(h)&={\bvarphi}_{-,p,s,a}\{1+o(1)\}, \qquad &\widetilde{\bvarphi}_{+,p,s,a}(h)&={\bvarphi}_{+,p,s,a}\{1+o(1)\},
\end{alignat*}
where
\begin{alignat*}{3}
     {\bzeta}_{-,p,s,a} &= f(c)\int\limits_{-\infty}^0 K(u)\br_{p,s}(u,\bmu_{W})u^{a+1} \de u, &{\bzeta}_{+,p,s,a}&=f(c)\int\limits_0^\infty K(u)\br_{p,s}(u,\bmu_{W})u^{a+1} \de u, \\
     {\bvarphi}_{-,p,s,a} &= f(c)\int\limits_{-\infty}^0 K(u) \begin{bmatrix}
    \br_p(u)\bmu_{W}^\prime \\
    \bmu_{WW}\otimes \br_s(u)
\end{bmatrix} u^{a+1}  \de u, &\quad  
     {\bvarphi}_{+,p,s,a}&=f(c)\int\limits_0^\infty K(u)\begin{bmatrix}
    \br_p(u)\bmu_{W}^\prime \\
    \bmu_{WW}\otimes \br_s(u)
\end{bmatrix}u^{a+1} \de u.
\end{alignat*}

\end{lemma}
\begin{flushright}
[\hyperref[proof: vartheta hat order]{Proof}]
\end{flushright}

\section{Main Results}\label{sec: main results}

In this section, we mostly focus on the RD estimator to the right of the cutoff as everything holds symmetrically for the RD estimator to the left. All the lemmas and main results are reported for both estimators.

The RD estimator with interacted covariates to the right of the cutoff is defined as 
\[\widehat{\bvartheta}_{+,p,s}(h):= \argmin_{\bt \in \R^{1+p+m}}\sum_{i=1}^n\indic(X_i\geq c)K\left(\frac{X_i-c}{h}\right)(Y_i - \br_{p,s}(X_i-c,\bW_i)^\prime\bt)^2.\]

To analyze the statistical properties of the RD estimator with interacted covariates $\widehat{\bvartheta}_{+,p,s}(h)$, we take advantage of the following decomposition:
\begin{align*}
    \widehat{\bvartheta}_{+,p,s}(h)&= \bH_{p,s}^{-1}(h)\widehat{\bGamma}_{+,p,s}^{-1}(h) \frac{1}{nh}\sum_{i=1}^n\indic(X_i\geq c)K\left(\frac{X_i-c}{h}\right)\br_{p,s}\left(\frac{X_i-c}{h},\bW_i\right)Y_i, \\
    &= \bvartheta_{+,p,s}^\star(h) +  \bH_{p,s}^{-1}(h)\widehat{\bGamma}_{+,p,s}^{-1}(h) \frac{1}{nh}\sum_{i=1}^n\indic(X_i\geq c)K\left(\frac{X_i-c}{h}\right)\br_{p,s}\left(\frac{X_i-c}{h},\bW_i\right)u_{+,i}(h),
\end{align*}
where $\bvartheta_{+,p,s}^\star(h)$ the fixed-$h$ best linear mean square error predictor of $\E[Y_i(1)\mid X_i=c, \bW_i]$ and
\[u_{+,i}(h):= Y_i -  \br_{p,s}\left(X_i-c,\bW_i\right)^\prime \bvartheta^\star_{+,p,s}(h)\]
is the fixed-$h$ $\ell_2$-projection residual. This decomposition gives us
\begin{align*}
    \widehat{\bvartheta}_{+,p,s}(h)- \bvartheta^\star_{+,p,s}(h) = \bH_{p,s}^{-1}(h)\widehat{\bGamma}_{+,p,s}^{-1}(h)\bL_{+,p,s}(h),
\end{align*}
where
\begin{align*}
    \bL_{+,p,s}(h)&=\frac{1}{nh}\sum_{i=1}^n\indic(X_i\geq c)K\left(\frac{X_i-c}{h}\right)\br_{p,s}\left(\frac{X_i-c}{h},\bW_i\right)u_{+,i}(h).
\end{align*}

\subsection{Asymptotic Approximation and Asymptotic Variance}

Lemma \ref{lemma: grand hessian} shows that $\widehat{\bGamma}_{+,p,s}(h)$ is asymptotically invertible. This Lemma takes care of the ``denominator" of the RD estimator by showing that
\[\widehat{\bGamma}_{+,p,s}(h) = {\bGamma}_{+,p,s} + o(1) + \Op(1/\sqrt{nh}).\]

Coming to the $\bL_{+,p,s}(h)$ term,
first note that $\E[\bL_{+,p,s}(h)] = 0$ because of the properties of $\ell_2$-projection residuals. Regarding the variance of this term,  the next lemma shows that $\V[\bL_{+,p,s}(h)]$ is $O(1/\sqrt{nh})$. Then, it follows that $\bL_{+,p,s}(h) = \Op(1/\sqrt{nh})$ by Lemma \ref{lemma: markov}.

\begin{lemma}
\label{lemma: first order variance}
    Let Assumptions \ref{ass: sampling}, \ref{ass: density running variable}, \ref{ass: DGP}, and \ref{ass: kernel} hold with $\mathsf{k} \geq h$. Then, for fixed $h>0$
    \begin{align*}
        \V[\bL_{-,p,s}(h)] = \frac{1}{nh}\mathbf{V}_{-,p,s}(h), \qquad \V[\bL_{+,p,s}(h)] = \frac{1}{nh}\mathbf{V}_{+,p,s}(h).
    \end{align*}
    Furthermore, as $nh\to\infty$ and $h\to 0$ 
    \begin{alignat*}{3}
        \mathbf{V}_{-,p,s}(h) &=\mathbf{V}_{-,p,s}\{1+o(1)\}, \quad &\mathbf{V}_{+,p,s}(h) &=\mathbf{V}_{+,p,s}\{1+o(1)\},
    \end{alignat*}
where $|\mathbf{V}_{-,p,s}|<\infty$ and $|\mathbf{V}_{+,p,s}|<\infty$.
\end{lemma}
\begin{flushright}
[\hyperref[proof: first order variance]{Proof}]
\end{flushright}

Typically, we are interested in estimators of the form $\cbe_{\nu}^\prime \widehat{\bvartheta}_{+,p,s}(h_+)$. By Lemma \ref{lemma: grand hessian} and Lemma \ref{lemma: first order variance} and Slutsky's theorem, we know that 
\[\sqrt{nh^{2\nu+1}}\cbe_{\nu}^\prime(\widehat{\bvartheta}_{+,p,s}(h) - \bvartheta_{+,p,s}^\star(h)) = \sqrt{nh}\cbe_{\nu}^\prime \widehat{\bGamma}_{+,p,s}^{-1}(h)\bL_{+,p,s}(h)=\Op(1).\]
The next theorem shows that the term above is not only bounded in probability but also converging in distribution to a Normal random variable. 
\begin{theorem}\label{thm: asymptotic distribution}
   Let Assumptions \ref{ass: sampling}, \ref{ass: density running variable}, \ref{ass: DGP}, and \ref{ass: kernel} hold with $\mathsf{k} \geq h$. If $nh\to \infty $ and $ h\to 0$, then 
  \begin{align*}
    \sqrt{nh}\bH_{p,s}(h)(\widehat{\bvartheta}_{-,p,s}(h) - \bvartheta^\star_{-,p,s}(h)) &\cdist \mathsf{N}(\mathbf{0}_{1+p+d},\bOmega_{-,p,s}), \\
    \sqrt{nh}\bH_{p,s}(h)(\widehat{\bvartheta}_{+,p,s}(h) - \bvartheta^\star_{+,p,s}(h)) &\cdist \mathsf{N}(\mathbf{0}_{1+p+d},\bOmega_{+,p,s}),
   \end{align*}
   with
   \begin{align*}
    \bOmega_{-,p,s}&={\bGamma}_{-,p,s}^{-1}\mathbf{V}_{-,p,s}
    {{\bGamma}_{-,p,s}^{-1\prime}}, \qquad
    \bOmega_{+,p,s}={\bGamma}_{+,p,s}^{-1}\mathbf{V}_{+,p,s}
    {{\bGamma}_{+,p,s}^{-1\prime}},
   \end{align*}
   where $\bOmega_{-,p,s}$ and $\bOmega_{+,p,s}$ are positive definite matrices.
\end{theorem}

\begin{flushright}
[\hyperref[proof: asymptotic distribution]{Proof}]
\end{flushright}

\begin{corollary}
    Let the conditions in Theorem \ref{thm: asymptotic distribution} hold, then for $\nu\in\{0,\cdots,p\wedge s\}$:
  \begin{align*}
    \sqrt{nh^{2\nu+1}}\cbe_\nu^\prime(\widehat{\bvartheta}_{-,p,s}(h) - \bvartheta^\star_{-,p,s}(h)) &\cdist \mathsf{N}({0},\mathcal{V}_{-,\nu,p,s}), \\
    \sqrt{nh^{2\nu+1}}\cbe_\nu^\prime(\widehat{\bvartheta}_{+,p,s}(h) - \bvartheta^\star_{+,p,s}(h)) &\cdist \mathsf{N}({0},\mathcal{V}_{+,\nu,p,s}),
   \end{align*}
   where 
   \[\mathcal{V}_{-,\nu,p,s}:= \cbe_\nu^\prime \bOmega_{-,p,s}\cbe_\nu,\qquad\mathcal{V}_{+,\nu,p,s}:= \cbe_\nu^\prime \bOmega_{+,p,s}\cbe_\nu.\]
\end{corollary}

\subsection{Variance Estimation}

To estimate the asymptotic variance of the RD estimator, we propose the following \textit{plug-in} estimator
\[\widehat{\mathcal{V}}_{+,\nu,p,s}(h) = \cbe_{\nu}^\prime\widehat{\bGamma}_{+,p,s}^{-1}(h) \widehat{\mathbf{V}}_{+,p,s}(h) {\widehat{\bGamma}_{+,p,s}^{-1}(h)}^\prime\cbe_{\nu},\]
where 
\[\widehat{\mathbf{V}}_{+,p,s}(h):= \frac{1}{nh}\sum_{i=1}^n\sqrt{w_{+,i}(h)}\cdot\indic(X_i\geq c)K\left(\frac{X_i-c}{h}\right)\br_{p,s}\left(\frac{X_i-c}{h},\bW_i\right)\br_{p,s}\left(\frac{X_i-c}{h},\bW_i\right)^\prime {\widehat{u}_{+,i}}^2(h),\]
with $\widehat{u}_{+,i}(h):= Y_i - \br_{p,s}(X_i-c,\bW_i)^\prime\widehat{\bvartheta}_{+,p,s}(h)$.

The weights $w_{+,i}(h)$ allow for different HC-type estimators. Namely,
\begin{table}[!ht]\centering
\begin{tabular}{ccccc}
\toprule & $\mathrm{HC}_0$ & $\mathrm{HC}_1$ & $\mathrm{HC}_2$ & $\mathrm{HC}_3$ \\
\midrule$w_{-, i}(h)$ & 1 & $\frac{N_{-}}{N_{-}-2 \operatorname{tr}\left(\mathbf{Q}_{-, p,s}\right)+\operatorname{tr}\left(\mathbf{Q}_{-, p,s} \mathbf{Q}_{-, p,s}\right)}$ & $\frac{1}{1-L_{+,i}}$ & $\frac{1}{\left(1-L_{+,i}\right)^2}$ \\
$w_{+, i}(h)$ & 1 & $\frac{N_{+}}{N_{+}-2 \operatorname{tr}\left(\mathbf{Q}_{+, p,s}\right)+\operatorname{tr}\left(\mathbf{Q}_{+, p,s} \mathbf{Q}_{+, p,s}\right)}$ & $\frac{1}{1-L_{+,i}}$ & $\frac{1}{\left(1-L_{+,i}\right)^2}$ \\
 \bottomrule
\end{tabular}
\end{table}

where
\begin{align*}
N_{-}:=\sum_{i=1}^n \indic\left(X_i<c\right) \quad \text { and } \quad N_{+}:=\sum_{i=1}^n \indic\left(X_i \geq c\right),
\end{align*}
$\mathbf{Q}_{-, p,s}, $ and $ \mathbf{Q}_{+, p,s}$ are the ``projection" matrices used to get the estimated residuals and are defined as,
\begin{align*}
\mathbf{Q}_{-, p,s}:=\mathbf{R}_{p,s}(h) \widehat{\bGamma}_{-, p,s}^{-1} \mathbf{R}_{p,s}(h)^{\prime} \mathbf{K}_{-}(h) / n, \quad \mathbf{Q}_{+, p,s}=\mathbf{R}_{p,s}(h) \widehat{\bGamma}_{+, p,s}^{-1} \mathbf{R}_{p,s}(h)^{\prime} \mathbf{K}_{+}(h) / n,
\end{align*}
and
\[L_{-,i} := \mathbf{e}_i^{\prime} \mathbf{Q}_{-, p,s} \mathbf{e}_i,\qquad L_{+,i} := \mathbf{e}_i^{\prime} \mathbf{Q}_{+, p,s} \mathbf{e}_i,\]
which denote the leverage of each observation.
\begin{theorem}
    \label{theorem: variance estimation plug-in}
     Let the assumptions of Theorem \ref{thm: asymptotic distribution} hold, then
    \begin{align*}
        \widehat{\mathcal{V}}_{-,\nu,p,s}(h) \cprob {\mathcal{V}}_{-,\nu,p,s}, \qquad \widehat{\mathcal{V}}_{+,\nu,p,s}(h) \cprob{\mathcal{V}}_{+,\nu,p,s}.
    \end{align*}
\end{theorem}
\begin{flushright}
[\hyperref[proof: variance estimation plug-in]{Proof}]
\end{flushright}

Define the standard error of the RD estimators as
\[\widehat{\sigma}_{-,\nu,p,s}(h) := \left(\frac{1}{{nh^{2\nu+1}}}\widehat{\mathcal{V}}_{-,\nu,p,s}(h)\right)^{1/2}, \qquad \widehat{\sigma}_{+,\nu,p,s}(h) := \left(\frac{1}{{nh^{2\nu+1}}}\widehat{\mathcal{V}}_{+,\nu,p,s}(h)\right)^{1/2}.\]
The lemma above naturally yields the following corollary via Slutsky's theorem.

\begin{corollary}
Let Assumptions \ref{ass: sampling}, \ref{ass: density running variable}, \ref{ass: DGP}, and \ref{ass: kernel} hold with $\mathsf{k} \geq h$. If $nh^{2\nu+1}\to \infty$ and $h\to 0$ then for $\nu\in\{0,1,\ldots,p\wedge s\}$
   \begin{align*}
    \widehat{\sigma}_{-,\nu,p,s}^{-1}(h)\cbe_{\nu}^\prime(\widehat{\bvartheta}_{-,p,s}(h) - \bvartheta^\star_{-,p,s}(h)) &\cdist \mathsf{N}(0,1), \\
    \widehat{\sigma}_{+,\nu,p,s}^{-1}(h)\cbe_{\nu}^\prime(\widehat{\bvartheta}_{+,p,s}(h) - \bvartheta^\star_{+,p,s}(h)) &\cdist \mathsf{N}(0,1).
   \end{align*}    
\end{corollary}

\subsubsection*{Clustered Data}
In the case of clustered data, the extension of the above results is immediate. The only difference would be reflected in the form of the ``meat" matrices $\mathbf{V}_{-,\nu,p,s},$ and $ \mathbf{V}_{+,\nu,p,s}$, which will ultimately depend on the particular form of clustering being used. For a review on cluster-robust inference, see \cite{Cameron2015_JHR} and \cite{mackinnon2023ClusterrobustInferenceGuide}.

We assume that each unit $i\in\{1,2,\cdots,n\}$ belongs to a single cluster $s(i)\in\{1,2,\cdots,G\},$ where $s:\{1,2,\cdots,n\}\to\{1,2,\cdots,G\}$. Furthermore, we consider an asymptotic regime where the number of clusters $G$ grows large, $G\to\infty$, and dominates the bandwidth $Gh\to\infty$. 

To estimate the asymptotic variance of the RD estimator under clustering, we propose the following \textit{plug-in} estimator
\[\widehat{\mathcal{V}}^{\mathsf{CL}}_{+,\nu,p,s}(h) = \frac{(G-1)n}{(G-1)(n-p-1-d)}\cbe_{\nu}^\prime\widehat{\bGamma}_{+,p,s}^{-1}(h) \widehat{\mathbf{V}}^{\mathsf{CL}}_{+,p,s}(h) {\widehat{\bGamma}_{+,p,s}^{-1}(h)}^\prime\cbe_{\nu},\]
where 
\begin{align*}
    \widehat{\mathbf{V}}^{\mathsf{CL}}_{+,p,s}(h)&:= \frac{1}{Gh}\sum_{g=1}^G\sum_{i,j=1}^n\indic(s(i)=s(j))\cdot\indic(X_i\geq c, X_j\geq c)\times \\
    &\qquad\quad K\left(\frac{X_i-c}{h}\right)K\left(\frac{X_j-c}{h}\right)\br_{p,s}\left(\frac{X_i-c}{h},\bW_i\right)\br_{p,s}\left(\frac{X_j-c}{h},\bW_j\right)^\prime {\widehat{u}_{+,i}}(h){\widehat{u}_{+,j}}(h).
\end{align*}

This estimator, as well as many more, is implemented in our \texttt{R} and \texttt{Stata} software.

\subsection{Smoothing Bias and Probability Limit}

Without further assumptions on the data generating process or on $\mu_{Y-}(X_i,\bW_i)$ and $\mu_{Y+}(X_i,\bW_i)$, the probability limit of the RD estimator has a generic best linear mean squared error predictor interpretation. We need to leverage extra structure to attach more interpretation to this probability limit. In this spirit, we posit Assumption \ref{ass: conditional expectation}, which we state again here below for the reader's convenience.
\addtocounter{assumption}{-2}
\begin{assumption}
The expectation of $Y_i(t), t\in\{0,1\}$ conditional on $(X_i,\bW_i)$ is
\begin{align*}
    \mu_-(X_i, \bW_i) = \alpha_{-}(X_i) + \blambda_{-}(X_i)^\prime \bW_i, \qquad
    \mu_+(X_i, \bW_i) = \alpha_{+}(X_i) + \blambda_{+}(X_i)^\prime \bW_i,
\end{align*}
where $\blambda_{-}(x):=(\lambda_{-,1}(x), \cdots, \lambda_{-,d}(x))^\prime, \blambda_{+}(x):=(\lambda_{+,1}(x), \cdots, \lambda_{+,d}(x))^\prime,$ and $\alpha_{-}(x),\alpha_{+}(x)\in\mathcal{C}^{q+2},$ and  $\lambda_{-,\ell}(x),\lambda_{+,\ell}(x),\in\mathcal{C}^{q+2}$ for all $\ell\in\{1,\cdots,d\}$ for some $q\in\N$.     
\end{assumption}

Let
\begin{align}\label{eq: xi star under partial linear model}
    \bvartheta_{-,p,s} := \begin{bmatrix}
        \balpha_{-,p}(c) \\
        \blambda_{-,s}(c)
    \end{bmatrix}, \qquad
    \bvartheta_{+,p,s} := \begin{bmatrix}
        \balpha_{+,p}(c) \\
        \blambda_{+,s}(c)
    \end{bmatrix},
\end{align}
where
\begin{align*}
    \balpha_{-,p}(x) &= \begin{bmatrix}
        \alpha_{-}(x) \\ \alpha_{-}^{(1)}(x) \\ \vdots \\
        \frac{\alpha_{-}^{(p)}(x)}{p!}
    \end{bmatrix}, \quad &\blambda_{-,s}(x) &=
    \begin{bmatrix}
        \lambda_{-,1}(x) \\ \vdots \\
        \frac{\lambda_{-,1}^{(s)}(x)}{s!} \\ \vdots \\
        \lambda_{-,d}(x) \\ \vdots \\
        \frac{\lambda_{-,d}^{(s)}(x)}{s!}
    \end{bmatrix}, \quad
    \balpha_{+,p}(x) &= \begin{bmatrix}
        \alpha_{+}(x) \\ \alpha_{+}^{(1)}(x) \\ \vdots \\
        \frac{\alpha_{+}^{(p)}(x)}{p!}
    \end{bmatrix}, \quad &\blambda_{+,s}(x) &=
    \begin{bmatrix}
        \lambda_{+,1}(x) \\ \vdots \\
        \frac{\lambda_{+,1}^{(s)}(x)}{s!} \\ \vdots \\
        \lambda_{+,d}(x) \\ \vdots \\
        \frac{\lambda_{+,d}^{(s)}(x)}{s!}
    \end{bmatrix},
\end{align*}
with $\balpha_{-,p}\equiv \balpha_{-,p}(c), \blambda_{-,s}\equiv \blambda_{-,s}(c),\balpha_{+,p}\equiv \balpha_{+,p}(c),$ and $ \blambda_{+,s}\equiv \blambda_{+,s}(c)$. 

Consider the following decomposition
\begin{align*}
    \bH_{p,s}(h)(\widehat{\bvartheta}_{+,p,s}(h) - \bvartheta_{+,p,s}) = \bH_{p,s}(h)(\widehat{\bvartheta}_{+,p,s}(h) - {\bvartheta}^\star_{+,p,s}(h)) + \bH_{p,s}(h)({\bvartheta}^\star_{+,p,s}(h) - \bvartheta_{+,p,s}).
\end{align*}
From Theorem \ref{thm: asymptotic distribution}, we know that $\bH_{p,s}(h)(\widehat{\bvartheta}_{+,p,s}(h) - {\bvartheta}^\star_{+,p,s}(h)) = \Op(1/\sqrt{nh})$. In what follows, we show that the second term is of order $\Op(h^{1+p\wedge s})$. Throughout, we consider an asymptotic regime in which this term is negligible with respect to the first term. It is precisely in this spirit that we will refer to ${\bvartheta}^\star_{+,p,s}(h) - \bvartheta_{+,p,s}$ as a ``bias" term. The next lemma shows that the leading term of the bias is of order $O(h^{1+p\wedge s})$.

\begin{lemma}[Smoothing Bias]\label{lemma: smoothing bias}
    Let Assumptions \ref{ass: sampling}-\ref{ass: kernel} hold for some $q\geq 2 + p\vee s$ and with $\mathsf{k} \geq h$. If $nh\to 0$ and $h\to 0$, then
    \begin{align*}
        {\bvartheta}_{-,p,s}^\star(h) &- \bvartheta_{-,p,s} =\\
        &\bH_{p,s}^{-1}(h)\left(h^{1+p}\widetilde{\bB}^{[0]}_{-,p}(h) + h^{2+p}\widetilde{\bB}^{[0]}_{-,p+1}(h) + h^{1+s}\widetilde{\bB}^{[1]}_{-,s}(h) + h^{2+s}\widetilde{\bB}^{[1]}_{-,s+1}(h) + o(h^{2+p\wedge s})\right),\\
        {\bvartheta}_{+,p,s}^\star(h) &- \bvartheta_{+,p,s} =\\
        &\bH_{p,s}^{-1}(h)\left(h^{1+p}\widetilde{\bB}^{[0]}_{+,p}(h) + h^{2+p}\widetilde{\bB}^{[0]}_{+,p+1}(h) + h^{1+s}\widetilde{\bB}^{[1]}_{+,s}(h) + h^{2+s}\widetilde{\bB}^{[1]}_{+,s+1}(h) + o(h^{2+p\wedge s})\right),\\
    \end{align*}
    where
    \begin{alignat*}{3}
        \widetilde{\bB}^{[0]}_{-,a}(h) =\widetilde{\bGamma}_{-,p,s}^{-1}(h)\widetilde{\bzeta}_{-,p,s,a}(h) \frac{\alpha^{(a+1)}_{-}(c)}{(a+1)!}\to {\bGamma}_{-,p,s}^{-1}{\bzeta}_{-,p,s,a} \frac{\alpha^{(a+1)}_{+}(c)}{(a+1)!}=:\bB^{[0]}_{-,a}, \\  
        \widetilde{\bB}^{[0]}_{+,a}(h) =\widetilde{\bGamma}_{+,p,s}^{-1}(h)\widetilde{\bzeta}_{+,p,s,a}(h) \frac{\alpha^{(a+1)}_{+}(c)}{(a+1)!}\to {\bGamma}_{+,p,s}^{-1}{\bzeta}_{+,p,s,a} \frac{\alpha^{(a+1)}_{+}(c)}{(a+1)!}=:\bB^{[0]}_{+,a}, \\  
        \widetilde{\bB}^{[1]}_{-,a}(h) = \widetilde{\bGamma}_{-,p,s}^{-1}(h)\widetilde{\bvarphi}_{-,p,s,a}(h) \frac{\blambda^{(a+1)}_{-}(c)}{(a+1)!}\to {\bGamma}_{-,p,s}^{-1}{\bvarphi}_{-,p,s,a}\frac{\blambda^{(a+1)}_{-}(c)}{(a+1)!}=:\bB^{[1]}_{-,a}, \\
        \widetilde{\bB}^{[1]}_{+,a}(h) = \widetilde{\bGamma}_{+,p,s}^{-1}(h)\widetilde{\bvarphi}_{+,p,s,a}(h) \frac{\blambda^{(a+1)}_{+}(c)}{(a+1)!}\to {\bGamma}_{+,p,s}^{-1}{\bvarphi}_{+,p,s,a}\frac{\blambda^{(a+1)}_{+}(c)}{(a+1)!}=:\bB^{[1]}_{+,a}.
    \end{alignat*}
\end{lemma}
\begin{flushright}
[\hyperref[proof: smoothing bias]{Proof}]
\end{flushright}

The following corollary follows naturally from Theorem \ref{thm: asymptotic distribution} and Lemma \ref{lemma: smoothing bias}.
\begin{corollary}\label{cor: asympt distr under sa5}
Let the assumptions of Lemma \ref{lemma: smoothing bias} hold. If $nh^{2\nu+1}\to \infty, h\to 0,$ and $nh^{2(p\wedge s) + 3}\to 0$, then 
   \begin{align*}
    \sqrt{nh^{2\nu+1}}\cbe_{\nu}^\prime(\widehat{\bvartheta}_{-,p,s}(h) - \bvartheta_{-,p,s})&\cdist \mathsf{N}(0,\mathcal{V}_{-,\nu,p,s}), \\
    \sqrt{nh^{2\nu+1}}\cbe_{\nu}^\prime(\widehat{\bvartheta}_{+,p,s}(h) - \bvartheta_{+,p,s})&\cdist \mathsf{N}(0,\mathcal{V}_{+,\nu,p,s}),
   \end{align*}    
   where
   \[\mathcal{V}_{-,\nu,p,s}:= \cbe_\nu^\prime \bOmega_{-,p,s}\cbe_\nu,\qquad\mathcal{V}_{+,\nu,p,s}:= \cbe_\nu^\prime \bOmega_{+,p,s}\cbe_\nu,\]
    with
\begin{align*}
    \bOmega_{-,p,s}&={\bGamma}_{-,p,s}^{-1}\mathbf{V}_{-,p,s}
    {{\bGamma}_{-,p,s}^{-1\prime}}, \qquad
    \bOmega_{+,p,s}={\bGamma}_{+,p,s}^{-1}\mathbf{V}_{+,p,s} 
    {{\bGamma}_{+,p,s}^{-1\prime}},
   \end{align*}
   and where
    \small
    \begin{align*}
        \mathbf{V}_{-,p,s} &= f(c) \int\limits_{-\infty}^0 K^2(u)\begin{bmatrix}
            \br_p\left(u\right)\br_p\left(u\right)^\prime \E[\sigma^2_{Y-}(X_i,\bW_i)\mid X_i=c] & \E[\bW_i^\prime\sigma^2_{Y-}(X_i,\bW_i)\mid X_i=c]  \otimes \br_p\left(u\right)\br_s\left(u\right)^\prime \\
            \E[\bW_i \sigma^2_{Y-}(X_i,\bW_i)\mid X_i=c]  \otimes \br_s\left(u\right)\br_p\left(u\right)^\prime & \E[\bW_i\bW_i^\prime\sigma^2_{Y-}(X_i,\bW_i)\mid X_i=c]  \otimes \br_s\left(u\right)\br_s\left(u\right)^\prime
        \end{bmatrix} \de u,\\
        \mathbf{V}_{+,p,s} &= f(c) \int\limits_0^\infty K^2(u)\begin{bmatrix}
            \br_p\left(u\right)\br_p\left(u\right)^\prime \E[\sigma^2_{Y+}(X_i,\bW_i)\mid X_i=c] & \E[\bW_i^\prime\sigma^2_{Y+}(X_i,\bW_i)\mid X_i=c]  \otimes \br_p\left(u\right)\br_s\left(u\right)^\prime \\
            \E[\bW_i \sigma^2_{Y+}(X_i,\bW_i)\mid X_i=c]  \otimes \br_s\left(u\right)\br_p\left(u\right)^\prime & \E[\bW_i\bW_i^\prime\sigma^2_{Y+}(X_i,\bW_i)\mid X_i=c]  \otimes \br_s\left(u\right)\br_s\left(u\right)^\prime
        \end{bmatrix} \de u
    \end{align*}
    are positive-definite matrices.
\end{corollary}

\subsection{Consistency} \label{subsec: consistency}

Using Lemma \ref{lemma: first order variance} and Lemma  \ref{lemma: smoothing bias}, we have
\begin{align*}
    \bH_{p,s}(h)\left(\widehat{\bvartheta}_{-,p,s}(h) - \bvartheta_{-,p,s}\right) &= \Op\left(\frac{1}{\sqrt{nh}} + h^{1+p\wedge s}\right),\\
    \bH_{p,s}(h)\left(\widehat{\bvartheta}_{+,p,s}(h) - \bvartheta_{+,p,s}\right) 
    &= \Op\left(\frac{1}{\sqrt{nh}} + h^{1+p\wedge s}\right),
\end{align*}
thus if $nh \to \infty$ and $h\to 0$, then $\widehat{\bvartheta}_{-,p,s}(h) \cprob \bvartheta_{-,p,s}$ and $\widehat{\bvartheta}_{+,p,s}(h) \cprob \bvartheta_{+,p,s}$.

Typically, we are interested in estimators of the form $\cbe_{\nu}^\prime \widehat{\bvartheta}_{-,p,s}(h)$ or $\cbe_{\nu}^\prime \widehat{\bvartheta}_{+,p,s}(h)$. In this case, we get
\begin{align*}
    \cbe_{\nu}^\prime\left(\widehat{\bvartheta}_{-,p,s}(h) - \bvartheta_{-,p,s}\right) = \Op\left(\frac{1}{\sqrt{nh^{2\nu+1}}} + h^{1+p\wedge s-\nu}\right),\\
    \cbe_{\nu}^\prime\left(\widehat{\bvartheta}_{+,p,s}(h) - \bvartheta_{+,p,s}\right) = \Op\left(\frac{1}{\sqrt{nh^{2\nu+1}}} + h^{1+p\wedge s-\nu}\right).
\end{align*}
In an asymptotic regime where also $n\min\{h_-,h_+\}^{2\nu+1} \to \infty$ and $h\to 0$, then $\cbe_{\nu}^\prime\widehat{\bvartheta}_{-,p,s}(h) \cprob \cbe_{\nu}^\prime{\bvartheta}_{-,p,s}$ and $\cbe_{\nu}^\prime\widehat{\bvartheta}_{+,p,s}(h) \cprob \cbe_{\nu}^\prime{\bvartheta}_{+,p,s}$.

Define 
\[\widehat{\chi}_{\nu,p,s}(\bh) := \cbe_{\nu}^\prime\left(\widehat{\bvartheta}_{+,p,s}(h_+)-\widehat{\bvartheta}_{-,p,s}(h_-)\right), \quad {\chi}_{\nu,p,s} := \cbe_{\nu}^\prime\left({\bvartheta}_{+,p,s}-{\bvartheta}_{-,p,s}\right).\]
Then
\[\widehat{\chi}_{\nu,p,s}(\bh) - {\chi}_{\nu,p,s} =  \Op\left(\frac{1}{\sqrt{n\min\{h_-,h_+\}^{2\nu+1}}} + \max\{h_-,h_+\}^{1+p\wedge s-\nu}\right),\]
and $\widehat{\chi}_{\nu,p,s}(\bh) \cprob {\chi}_{\nu,p,s}.$

\subsection{Causal Interpretation of the Probability Limit}

Before delving into this section, we redefine the ``extractor" vector as
\[\cbe_{\nu_x,\nu_w} = \begin{bmatrix}
    \nu_x\bI_{p+1}\be_{\nu_x} \\ \nu_w!\biota_{d} \otimes \bI_{s+1}\be_{{\nu_w}}
\end{bmatrix}\]
to be the vector extracting the $(\nu_x+1)$-th term in the polynomial basis in $(X_i-c)$ and the $(\nu_w+1)$-th terms in the interaction between such basis and $\bW_i$. Furthermore, we define $\cbe_{\nu_x,\cdot} \equiv \bI_{1+p+d}\be_{\nu_x},$ to be the vector that extracts only the $(\nu_x+1)$-th term in the polynomial basis in $(X_i-c)$. We also define accordingly
\[\chi_{\nu_x,\nu_w,p,s} := \cbe_{\nu_x,\nu_w}^\prime \left({\bvartheta}_{+,p,s}-{\bvartheta}_{-,p,s}\right).\]

The quantity $\chi_{\nu_x,\nu_w,p,s}$ has a natural causal interpretation in many cases, which depends on the elements extracted:
\begin{enumerate}
    \item When $d=1$ and $W_i$ is binary, we have that 
    \[\chi_{\nu_x,\cdot,p,s} = \frac{\partial^{\nu_x} }{\partial x^{\nu_x}}\E[Y_i(1)-Y_i(0)\mid X_i=x, W_i=0] \Big|_{x=c},\]
    and 
    \[\chi_{\nu,\nu,p,s} = \frac{\partial^{\nu} }{\partial x^{\nu}}\left(\E[Y_i(1)-Y_i(0)\mid X_i=x, W_i=1]-\E[Y_i(1)-Y_i(0)\mid X_i=x, W_i=0]\right) \Big|_{x=c}.\]
    In other words, if a researcher is interested in the conditional average treatment effect (CATE) for the two sub-populations defined by $W_i\in\{0,1\},$ then the first element of $\widehat{\bvartheta}_{+,p,s}(h_+)-\widehat{\bvartheta}_{-,p,s}(h_-)$ identifies the CATE for the baseline ($W_i=0$) group, whereas the $(2+p)$-th element identifies the difference in CATEs between in the sub-population with $W_i=1$ and the baseline one.
    \item When there is more than one binary covariate, say $U_i$ and $V_i$, then $\bW_i=(U_i,V_i,U_i\cdot V_i)^\prime$  to correctly identify the CATEs in all the sub-populations.
    \item If $d=1$ and the covariate of interest $W_i$ is categorical (e.g., race) or multi-valued discrete (e.g., age) --i.e., takes on $J$ distinct values $(1,\ldots,J)$-- then the same interpretation offered above holds as long as the covariate is dummied out. In other words, it means that 
    $$\bW_i = (I_2,I_3,\ldots,I_J)^\prime, \qquad I_j:=\indic(W_i=j), j=2,\ldots, J .$$
    \item If $d=1$ and $W_i$ is a continuous covariate (e.g., parental income), then we still have
    \[\chi_{\nu_x,\cdot,p,s} = \frac{\partial^{\nu_x} }{\partial x^{\nu_x}}\E[Y_i(1)-Y_i(0)\mid X_i=x, W_i=0] \Big|_{x=c},\]
    but the interpretation of the coefficients on the interaction terms between $(X_i-c)$ and $W_i$ changes as follows
    \[\chi_{\nu,p,s} = \frac{\partial^{\nu_x} }{\partial x^{\nu_x}}\frac{\partial^{\nu_w} }{\partial w^{\nu_w}}\E[Y_i(1)-Y_i(0)\mid X_i=x, W_i=w] \Big|_{x=c},\]
    where the evaluation point of the derivative in the $W$-dimension needs not to be specified as this derivative is constant in the $W$-dimension by Assumption \ref{ass: conditional expectation}. The coefficient on the interaction terms simply captures the change in the CATEs due to a marginal change in $W_i$. This change is assumed to be linear in virtue of Assumption \ref{ass: conditional expectation}. 
    \end{enumerate}

\subsection{MSE Expansions} \label{subsec: mse expansions}

We now provide first-order expansions for the MSE of the RD estimator. This is crucial to obtain formulas for MSE-optimal bandwidths. We first start by giving a Nagar expansion of the unconditional MSE. Then, we provide an approximation for the bias and the variance of the RD estimator, where the approximation comes from disregarding higher-order terms.

First of all, recall that the RD estimator and the best linear mean square error predictor are 
\[\widehat{\bvartheta}_{+,p,s}(h) =\bH_{p,s}^{-1}(h)\widehat{\bGamma}^{-1}_{+,p,s}(h)\widehat{\bUpsilon}_{+,p,s}(h), \qquad \bvartheta^\star_{+,p,s}(h) =\bH_{p,s}^{-1}(h)\widetilde{\bGamma}^{-1}_{+,p,s}(h)\widetilde{\bUpsilon}_{+,p,s}(h).\]
Consider the following decomposition for fixed $\bw\in\R^d$
\begin{align*}
    \cbe_\nu'&\widehat{\bvartheta}_{+,p,s}(h) - \mu_+^{(\nu)}(c,\bw)
    \\
    &=h^{-\nu}\cbe_\nu'\widehat{\bGamma}^{-1}_{+,p,s}(h)\left(\widehat{\bUpsilon}_{+,p,s}(h) - \widehat{\bGamma}_{+,p,s}(h)\bH_{p,s}(h)\bvartheta_{+,p,s}^\star(h)\right) + \cbe_\nu'\bvartheta_{+,p,s}^\star(h) - \mu_+^{(\nu)}(c,\bw) \\
    &= \underbrace{h^{-\nu}\cbe_\nu'\widetilde{\bGamma}^{-1}_{+,p,s}(h)\bL_{+,p,s}(h)}_{\mathsf{L}_{+,\nu,p,s}(h)} + \underbrace{h^{-\nu}\cbe_\nu'\left(\widehat{\bGamma}^{-1}_{+,p,s}(h)-\widetilde{\bGamma}^{-1}_{+,p,s}(h)\right)\bL_{+,p,s}(h)}_{\mathsf{Q}_{+,\nu,p,s}(h)} + \underbrace{\cbe_\nu'\bvartheta_{+,p,s}^\star(h) - \mu_+^{(\nu)}(c,\bw)}_{\mathsf{B}_{+,\nu,p,s}(h)},
\end{align*}
where we refer to $\mathsf{L}_{+,\nu,p,s}(h)$ as the ``linear" term, to $\mathsf{Q}_{+,\nu,p,s}(h)$ as the ``quadratic" term, and to $\mathsf{B}_{+,\nu,p,s}(h)$ as the ``bias" term. By Theorem \ref{thm: asymptotic distribution} and Lemma \ref{lemma: grand hessian}, we know that $\mathsf{L}_{+,\nu,p,s}(h) = \Op(1/\sqrt{nh^{2\nu+1}})$ and $\mathsf{Q}_{+,\nu,p,s}(h) = \Op(1/\sqrt{n^2h^{2\nu+2}}) = \op(\mathsf{L}_{+,\nu,p,s}(h))$. Moreover, under Assumption \ref{ass: conditional expectation}, we have that $\mu_+^{(\nu)}(c,\bW_i)=\cbe'\bvartheta_{+,p,s}(h)$, thus by Lemma \ref{lemma: smoothing bias} we have that $\mathsf{B}_{+,\nu,p,s}(h) = O(h^{1+p\wedge s -\nu})$. In an asymptotic regime in which $nh^{2+p\wedge s}\to \infty$, we have that $\mathsf{Q}_{+,\nu,p,s}(h)=\op(\mathsf{B}_{+,\nu,p,s}(h))$. Therefore, under Assumptions \ref{ass: sampling}-\ref{ass: conditional expectation}, as $n\to\infty$ and $nh^{2+p\wedge s}\to\infty$ we have
\[\cbe_\nu'(\widehat{\bvartheta}_{+,p,s}(h) - \bvartheta_{+,p,s}(h)) = \mathsf{L}_{+,\nu,p,s}(h) + \mathsf{B}_{+,\nu,p,s}(h) + \op(\min\{\mathsf{L}_{+,\nu,p,s}(h),\mathsf{B}_{+,\nu,p,s}(h)\}).\]
With this result at hand, we define the first-order approximation of the mean squared error as
\[\mathsf{MSE}[\cbe_{\nu}^\prime (\widehat{\bvartheta}_{+,p,s}(h) -\bvartheta_{+,p,s})]= \E\left[(\mathsf{L}_{+,\nu,p,s}(h) + \mathsf{B}_{+,\nu,p,s}(h))^2\right] = \V[\mathsf{L}_{+,\nu,p,s}(h)] + \mathsf{B}_{+,\nu,p,s}(h)^2,\]
where the second equality follows from the fact that $ \mathsf{B}_{+,\nu,p,s}(h)$ is non-random and $\mathsf{L}_{+,\nu,p,s}(h)$ is unconditionally mean-zero. Similarly, we define
\begin{align*}
    \mathsf{MSE}[\cbe_{\nu}^\prime (\widehat{\bvartheta}_{-,p,s}(h) -\bvartheta_{-,p,s})]&= \V[\mathsf{L}_{-,\nu,p,s}(h)] + \mathsf{B}_{-,\nu,p,s}(h)^2
\end{align*}
for the left side of the cutoff, whereas for the difference, we define
\begin{align*}
    \mathsf{MSE}[\cbe_{\nu}^\prime (\widehat{\bvartheta}_{+,p,s}(h)-\widehat{\bvartheta}_{-,p,s}(h) &-({\bvartheta}_{+,p,s}(h) -\bvartheta_{-,p,s}))] \\
    &= \V[\mathsf{L}_{+,\nu,p,s}(h)] +\V[\mathsf{L}_{-,\nu,p,s}(h)] + (\mathsf{B}_{+,\nu,p,s}(h)-\mathsf{B}_{-,\nu,p,s}(h))^2.
\end{align*}

Under Assumptions \ref{ass: sampling}-\ref{ass: kernel} and relying on Lemma \ref{lemma: smoothing bias}, the bias approximations for the RD estimators are given by
\begin{align*}
    \mathsf{B}_{-,\nu,p,s}(h) = h^{1+p-\nu} \mathcal{B}^{[0]}_{-,\nu,p,s}(h) + h^{1+s-\nu} \mathcal{B}^{[1]}_{-,\nu,p,s}(h) + o(h^{1+(p\wedge s)-\nu}), \\
    \mathsf{B}_{+,\nu,p,s}(h) = h^{1+p-\nu} \mathcal{B}^{[0]}_{+,\nu,p,s}(h) + h^{1+s-\nu} \mathcal{B}^{[1]}_{+,\nu,p,s}(h) + o(h^{1+(p\wedge s)-\nu}),
\end{align*}
where
\begin{alignat*}{3}
    \mathcal{B}^{[0]}_{-,\nu,p,s}(h) =\cbe_{\nu}^\prime\bB_{-,p,s}^{[0]}(h)&\to \cbe_{\nu}^\prime\bB_{-,p,s}^{[0]}=:\mathcal{B}^{[0]}_{-,\nu,p,s}, \quad \mathcal{B}^{[1]}_{-,\nu,p,s}(h) = \cbe_{\nu}^\prime\bB_{-,p,s}^{[1]}(h)&\to \cbe_{\nu}^\prime\bB_{-,p,s}^{[1]}(h)=:\mathcal{B}^{[1]}_{-,\nu,p,s}, \\
    \mathcal{B}^{[0]}_{+,\nu,p,s}(h) =\cbe_{\nu}^\prime\bB_{+,p,s}^{[0]}(h) &\to \cbe_{\nu}^\prime\bB_{-,p,s}^{[0]} =:\mathcal{B}^{[0]}_{+,\nu,p,s}, \quad
    \mathcal{B}^{[1]}_{+,\nu,p,s}(h) = \cbe_{\nu}^\prime\bB_{+,p,s}^{[1]}(h)&\to \cbe_{\nu}^\prime\bB_{+,p,s}^{[1]}(h)=:\mathcal{B}^{[1]}_{+,\nu,p,s}.
\end{alignat*}
Furthermore, we define
\begin{alignat*}{3}
    \mathcal{B}_{-,\nu,p,s}(h) &:=\indic(p\leq s)\mathcal{B}^{[0]}_{-,\nu,p,s}(h)+\indic(p\geq s)\mathcal{B}^{[1]}_{-,\nu,p,s}(h),\qquad
    &\mathcal{B}_{-,\nu,p,s} &:=\indic(p\leq s)\mathcal{B}^{[0]}_{-,\nu,p,s}+\indic(p\geq s)\mathcal{B}^{[1]}_{-,\nu,p,s},\\
    \mathcal{B}_{+,\nu,p,s}(h) &:=\indic(p\leq s)\mathcal{B}^{[0]}_{+,\nu,p,s}(h)+\indic(p\geq s)\mathcal{B}^{[1]}_{+,\nu,p,s}(h), \qquad
    &\mathcal{B}_{+,\nu,p,s} &:=\indic(p\leq s)\mathcal{B}^{[0]}_{+,\nu,p,s}+\indic(p\geq s)\mathcal{B}^{[1]}_{+,\nu,p,s}.
\end{alignat*}

Under Assumptions \ref{ass: sampling}-\ref{ass: kernel} and relying on Corollary \ref{cor: asympt distr under sa5}
\begin{align*}
    \V[\mathsf{L}_{-,\nu,p,s}(h)] &= \frac{1}{nh^{2\nu+1}}\mathcal{V}_{-,\nu,p,s}(h), \qquad \mathcal{V}_{-,\nu,p,s}(h):=\cbe_{\nu}^\prime\widetilde{\bGamma}_{-,p,s}^{-1}(h)\mathbf{V}_{-,p,s}(h)
    {\widetilde{\bGamma}_{-,p,s}^{-1\prime}(h)}\cbe_{\nu},\\
    \V[\mathsf{L}_{+,\nu,p,s}(h)] &= \frac{1}{nh^{2\nu+1}}\mathcal{V}_{+,\nu,p,s}(h),\qquad \mathcal{V}_{+,\nu,p,s}(h):=\cbe_{\nu}^\prime\widetilde{\bGamma}_{+,p,s}^{-1}(h)\mathbf{V}_{+,p,s}(h)
    {\widetilde{\bGamma}_{+,p,s}^{-1\prime}(h)}\cbe_{\nu}.
\end{align*}
The variance approximations for the RD estimators are given by
\[\mathcal{V}_{-,\nu,p,s}(h)\to\mathcal{V}_{-,\nu,p,s}:=\cbe_{\nu}^\prime{\bGamma}_{-,p,s}^{-1}\mathbf{V}_{-,p,s}
    {{\bGamma}_{-,p,s}^{-1\prime}}\cbe_{\nu}, \qquad
    \mathcal{V}_{+,\nu,p,s}(h)\to\mathcal{V}_{+,\nu,p,s}:=\cbe_{\nu}^\prime{\bGamma}_{+,p,s}^{-1} \mathbf{V}_{+,p,s}
    {{\bGamma}_{+,p,s}^{-1\prime}}\cbe_{\nu}.\]
 
With these approximations, the MSE expansions of the RD estimators are given by
\begin{align*}
    \mathsf{MSE}[\cbe_{\nu}^\prime (\widehat{\bvartheta}_{-,p,s}(h) -\bvartheta_{-,p,s})] &= \frac{1}{nh^{2\nu+1}}\mathcal{V}_{-,\nu,p,s} + h^{2(1+p\wedge s-\nu)}\mathcal{B}^2_{-,\nu,p,s}, \\
    \mathsf{MSE}[\cbe_{\nu}^\prime (\widehat{\bvartheta}_{+,p,s}(h) -\bvartheta_{+,p,s})] &= \frac{1}{nh^{2\nu+1}}\mathcal{V}_{+,\nu,p,s} + h^{2(1+p\wedge s-\nu)}\mathcal{B}^2_{+,\nu,p,s}.
\end{align*}
Accordingly, we define an MSE-optimal bandwidth as the minimizer of the Nagar expansion of the unconditional MSE.

\paragraph{One-sided Optimal Bandwidths.} The MSE-optimal bandwidths are defined as
\begin{align*}
    h^\star_{-,\nu,p,s} &:= \argmin_{h>0} \left[\frac{1}{nh^{2\nu+1}}\mathcal{V}_{-,\nu,p,s} + h^{2((p\wedge s)+1-\nu)}\mathcal{B}_{-,\nu,p,s}^2\right], \\ 
    h^\star_{+,\nu,p,s} &:= \argmin_{h>0} \left[\frac{1}{nh^{2\nu+1}}\mathcal{V}_{+,\nu,p,s} + h^{2((p\wedge s)+1-\nu)}\mathcal{B}_{+,\nu,p,s}^2\right],
\end{align*}
so, under the additional assumption that $\mathcal{B}_{-,\nu,p,s}\neq0\neq \mathcal{B}_{+,\nu,p,s}$, we get
\begin{align*}
    h^\star_{-,\nu,p,s} =\left[\frac{1+2\nu}{2(1+(p\wedge s)-\nu)n}\frac{\mathcal{V}_{-,\nu,p,s}}{\mathcal{B}_{-,\nu,p,s}^2}\right]^{\frac{1}{3+2(p\wedge s)}}, \qquad h^\star_{+,\nu,p,s} =\left[\frac{1+2\nu}{2(1+(p\wedge s)-\nu)n}\frac{\mathcal{V}_{+,\nu,p,s}}{\mathcal{B}_{+,\nu,p,s}^2}\right]^{\frac{1}{3+2(p\wedge s)}}.
\end{align*}

\paragraph{Two-sided Optimal Bandwidth.} In this case a single bandwidth is chosen, so $h=h_+=h_-$ and the optimal bandwidth is defined as
\begin{align*}
    h^\star_{\nu,p,s} := \argmin_{h>0} \left[\frac{1}{nh^{2\nu+1}}(\mathcal{V}_{+,\nu,p,s}+\mathcal{V}_{-,\nu,p,s}) + h^{2((p\wedge s)+1-\nu)}(\mathcal{B}_{+,\nu,p,s}-\mathcal{B}_{-,\nu,p,s})^2\right], 
\end{align*}
so, under the additional assumption that $\mathcal{B}_{+,\nu,p,s}-\mathcal{B}_{-,\nu,p,s}\neq 0$, we get
\[h^\star_{\nu,p,s} =\left[\frac{1+2\nu}{2(1+(p\wedge s)-\nu)n}\frac{\mathcal{V}_{+,\nu,p,s}+\mathcal{V}_{-,\nu,p,s}}{(\mathcal{B}_{+,\nu,p,s}-\mathcal{B}_{-,\nu,p,s})^2}\right]^{\frac{1}{3+2(p\wedge s)}}.\]

\section{Empirical Practice Investigating Covariate-Heterogeneity}
\label{sec: empirical app}

To examine common empirical practices regarding heterogeneous effects, we sampled papers from AEA journals over the past decade. The papers are listed in Table \ref{table:papers} below, along with a summarization of how treatment effect heterogeneity was studied in each, including the empirical setting, the type of heterogeneity, and details on how estimation and inference were conducted. For full details of each study, please refer to the original work.

\begin{sidewaystable}[!ht]
    \renewcommand{\arraystretch}{1.25}
    \centering
    \caption{Empirical Practice for RD-HTE}
	\resizebox{\textwidth}{!}{\small
	\begin{tabular}{p{5cm} p{4.5cm} p{3.5cm} p{2cm}  |   p{0.6cm} p{0.6cm} p{0.6cm}   |   m{1.6cm} m{0.6cm}  m{0.6cm} m{1.2cm} |  m{0.7cm} m{0.6cm} }
\toprule\toprule
    &      &     &  & \multicolumn{3}{c}{\textit{Heterogeneity}} & \multicolumn{4}{c}{\textit{Estimation}}                      & \multicolumn{2}{c}{\textit{Std. Errs.}} \\
	\textit{Paper}   &                        \textit{Outcome}                           &                               \textit{Treatment} &     \textit{Running Var.}      & \textit{Disc.}      & \textit{Time}       & \textit{Cont.}      & \textit{Bandwidth}          & \textit{Local} & \textit{Joint} & \textit{Covs/FE} & \textit{Robust}      & \textit{Cluster}      \\ \midrule
	\cite{Adams}     & Healthcare   utilization                         & Financial Assistance           & Poverty level    &                  & \checkmark        &            & None               &       &       &              &             &                 \\
	\cite{Akhtari-Moreira-Trucco2022_AER}   & Municipal   Bureaucracy                           & Political turnover             & Vote Margin  & \checkmark              &            &            & MSE-HTE      & \checkmark   &       & \checkmark          &             & \checkmark             \\
	\cite{Asher}     & Economic   development                            & Road construction              & Population  & \checkmark              &            &            & MSE         & \checkmark   &       & \checkmark          &             & \checkmark             \\
	\cite{Brollo}    & Political   corruption, quality & Government revenues            & Population  & \checkmark              &            & \checkmark        & None                     & \checkmark   & \checkmark          &             & \checkmark             \\
	\cite{Dell}      & Drug-related   violence                           & Drug enforcement               & Vote Margin  & \checkmark              & \checkmark        &            & Manual              & \checkmark   & \checkmark   & \checkmark          &             & \checkmark             \\
	\cite{Miralles}  & Savings,   Retirement behavior                    & Public pension benefits        & Age              & \checkmark              & \checkmark        &            & MSE         & \checkmark   &       & \checkmark          & \checkmark         &                 \\
	\cite{Han}       & Healthcare Utilization                    & Patient Cost-Sharing           & Age              & \checkmark              &            &            & Manual              & \checkmark   &       &              &             & \checkmark             \\
	\cite{Huh}       & Mortality,   risky behaviors                      & Teenage Driving                & Age              & \checkmark              &            &            & MSE         & \checkmark   &       &              & \checkmark         &                 \\
	\cite{Jones}     & Irrigation   Adoption, Profits                    & Access to water                & Spatial    & \checkmark              &            &            & Manual             & \checkmark   &       & \checkmark          &             & \checkmark             \\
	\cite{Lindo}     & Academic   Performance                            & Academic probation             & Test Scores      & \checkmark              &            &            & Manual             & \checkmark   &       & \checkmark          &             & \checkmark             \\
	\cite{McEwan}    & Economics Major   Choice                          & Higher letter grade            & Test Scores      & \checkmark              &            &            & MSE        & \checkmark   & \checkmark   & \checkmark          & \checkmark         &                 \\
	\cite{Miglino}   & Health outcomes                                   & Financial Assistance           & Age              & \checkmark              & \checkmark        &            & Manual             & \checkmark   &       & \checkmark          &             & \checkmark             \\
	\cite{Pop-Eleches-Urquiola2013_AER}   & Academic   Performance                            & Access to better schools       & Test Scores      & \checkmark              & \checkmark        &            & Manual             & \checkmark   &       & \checkmark          &             & \checkmark             \\
	\cite{Shigeoka}  & Utilization,   health            & Patient Cost Sharing           & Age              & \checkmark              & \checkmark        &            & Manual            & \checkmark   &       & \checkmark          &             & \checkmark             \\
	\cite{Silliman}  & Labor market   returns                            & Vocational Education & Test Scores      & \checkmark              & \checkmark        &            & Manual           & \checkmark   &       & \checkmark          &             & \checkmark             \\
	\cite{Zimmerman} & Better jobs,   Income                             & Access to better schools       & Test Scores      & \checkmark              &            &            & Manual             & \checkmark   & \checkmark   & \checkmark          &             & \checkmark       \\     \bottomrule\bottomrule
\end{tabular}
    }
    \footnotesize{
    \begin{flushleft}Notes:\end{flushleft}\vspace{-0.2in}
    \begin{enumerate}[label=\normalfont(\roman*),noitemsep,leftmargin=*]
        \item The left four columns give the citation and the empirical setting in terms of the outcome variable, treatment studied, and running variable. 
        
        \item The type of heterogeneity is studied is listed in columns 5--7, showing, respectively, if the covariates used for heterogeneity are discrete (``Disc.''), time-related (``Time''), or continuous (``Cont.''). 

        \item For estimation, the ``Bandwidth'' column reports the method of bandwidth selection used, where ``Manual'' indicates a hand-selected bandwidth, ``MSE'' for a mean squared error optimal bandwidth, and ``MSE-HTE'' if the bandwidth was chosen particularly for the heterogeneity analysis. The ``Local'' indicates that the analysis is conducted local to the cutoff; ``Joint'' if the heterogeneity was conducted jointly across levels of the variable, and ``Covs/FE'' if additional covariates or fixed effects are included for efficiency. 

        \item Standard errors are described in the final two columns, indicating if robust standard errors are used or if the standard errors are clustered. 
    \end{enumerate}
    }
\label{table:papers}
\end{sidewaystable}

\cleardoublepage 
\section{Proofs}
\label{sec: proofs}

\subsection{Proof of Lemma \ref{lemma: markov}}
\begin{proof}\label{proof: markov}
Here, for simplicity, we just prove the lemma for the scalar case. By Markov inequality, we have 
\[\forall\, M>0,\quad \PP(|A_n - \E[A_n]| \geq M) \leq \frac{\E[|A_n-\E[A_n]|]}{M} \leq \frac{\V[A_n]^{1/2}}{M},\]
which implies that $\V[A_n]^{-1/2}(A_n - \E[A_n]) = \Op(1)$ which was to be shown. The matrix case follows using $\PP(\bA_n\preceq \bI ) \leq \mathrm{tr}(\E[\bA_n])$, where $\bA_n\preceq \bI$ means that $\bA_n- \bI$ is negative semi-definite.
\end{proof}

\subsection{Proof of Lemma \ref{lemma: hessian}}
\begin{proof}\label{proof: hessian}
The proof covers the case to the right of the cutoff. Everything follows symmetrically for the other case. First, for any $p,s\in\N$
\begin{align*}
    \E[\widehat{\bLambda}_{+,p,s}(h)] &= \E\left[\frac{1}{n}\sum_{i=1}^n\indic ( X_i\geq c)\br_p\left(\frac{X_i-c}{h}\right)\br_s\left(\frac{X_i-c}{h}\right)^\prime K\left(\frac{X_i-c}{h}\right)\frac{1}{h}\right] \tag{definition}\\
    &=\frac{1}{h}\E\left[\indic ( X_i\geq c)\br_p\left(\frac{X_i-c}{h}\right)\br_s\left(\frac{X_i-c}{h}\right)^\prime K\left(\frac{X_i-c}{h}\right)\right] \tag{Assumption \ref{ass: sampling}} \\
    &= \frac{1}{h}\int\limits_{c}^\infty \br_p\left(\frac{x-c}{h}\right)\br_s\left(\frac{x-c}{h}\right)^\prime K\left(\frac{x-c}{h}\right)f(x)\de x \\
    &= \int\limits_0^\infty \br_p(u)\br_s(u)^\prime K(u)f(hu+c) \de u \tag{$u = (x-c)/h$} 
    =: \widetilde{\bGamma}_{+,p,s}(h).
\end{align*}

By Lemma \ref{lemma: markov}
\[\widehat{\bLambda}_{+,p,s}(h) = \E[\widehat{\bLambda}_{+,p,s}(h)] + \Op(|\V[\widehat{\bLambda}_{+,p,s}(h)]|).\]
The generic element of $\widehat{\bLambda}_{+,p,s}(h)$ is
\[\frac{1}{nh}\sum_{i=1}^n\indic(X_i\geq c)\left(\frac{X_i-c}{h}\right)^{j} K\left(\frac{X_i-c}{h}\right), \quad j\in\{0,\cdots,p+s\}.\]
Fix $j\in\{0,\cdots,p+s\}$, then 
\begin{align*}
&\V\left[\frac{1}{nh}\sum_{i=1}^n\indic ( X_i\geq c)\left(\frac{X_i-c}{h}\right)^{j} K\left(\frac{X_i-c}{h}\right)\right] \\
&\qquad\qquad= \frac{1}{nh^2}\V\left[\indic ( X_i\geq c)\left(\frac{X_i-c}{h}\right)^{j} K\left(\frac{X_i-c}{h}\right)\right] \tag{Assumption \ref{ass: sampling}} \\
&\qquad \qquad \leq \frac{1}{nh^2}\E\left[\indic ( X_i\geq c)\left(\frac{X_i-c}{h}\right)^{2j} K\left(\frac{X_i-c}{h}\right)^2\right]\\
&\qquad \qquad = \frac{1}{nh}\int\limits_0^\infty u^{2j} K(u)^2 f(uh+c)\de u = O(n^{-1}h^{-1}), \tag{$u = (x-c)/h$} 
\end{align*}
where the last equality follows because $u^{2j} K(u)^2 f(uh+c)$ is integrable due to Assumptions \ref{ass: density running variable}, \ref{ass: kernel}, and $\mathsf{k}\geq h$. Hence $|\V[\widehat{\bLambda}_{+,p,s}(h)]| = O(n^{-1}h^{-1})$. Finally, we have $\widehat{\bLambda}_{+,p,s}(h) = \E[\widehat{\bLambda}_{+,p,s}(h)] + \Op(1/\sqrt{nh})$, which was to be shown.
\end{proof}

\subsection{Proof of Lemma \ref{lemma: grand hessian}}
\begin{proof}\label{proof: grand hessian}
    The proof covers the case to the right of the cutoff. Throughout the proof, we maintain that $n$ is large enough so that $\mathsf{k}\leq c+h$ in  Assumption \ref{ass: DGP}. Everything follows symmetrically for the other case. First, note that
    \[ \widehat{\bGamma}_{+,p,s}(h) = \begin{bmatrix}
        \widehat{\bLambda}_{+,p,p}(h) & \widehat{\bG}_{+,2}(h) \\ \widehat{\bG}_{+,2}(h)^\prime & \widehat{\bG}_{+,3}(h) 
    \end{bmatrix} ,\]
    where
    \begin{align*}
        \widehat{\bLambda}_{+,p,p}(h) &= \frac{1}{nh}\sum_{i=1}^n\indic ( X_i\geq c) \br_p\left(\frac{X_i-c}{h}\right)\br_p\left(\frac{X_i-c}{h}\right)^\prime K\left(\frac{X_i-c}{h}\right), \\ 
        \widehat{\bG}_{+,2}(h)&=\frac{1}{nh}\sum_{i=1}^n\indic ( X_i\geq c) \left[\bW_i^\prime\otimes K\left(\frac{X_i-c}{h}\right)\br_p\left(\frac{X_i-c}{h}\right)\br_s\left(\frac{X_i-c}{h}\right)^\prime\right], \\
         \widehat{\bG}_{+,3}(h)&=\frac{1}{nh}\sum_{i=1}^n\indic ( X_i\geq c) \left[\bW_i\bW_i^\prime\otimes K\left(\frac{X_i-c}{h}\right)\br_s\left(\frac{X_i-c}{h}\right)\br_s\left(\frac{X_i-c}{h}\right)^\prime\right]. 
    \end{align*}
    Then, from Lemma \ref{lemma: hessian} we immediately have
    \[\widehat{\bLambda}_{+,p,p}(h) = \widetilde{\bLambda}_{+,p,p}(h) + \Op(1/\sqrt{nh}).\]
    
    Consider the expectation of $\widehat{\bG}_{+,2}(h)$. Using Assumption \ref{ass: sampling} and the change of variable $u = (x-c)/h$:
    \begin{align*}
        \E[\widehat{\bG}_{+,2}(h)] &= \E\left[\frac{1}{h}\indic ( X_i\geq c)\left\{\bmu_{W}(X_i)^\prime\otimes K\left(\frac{X_i-c}{h}\right)\br_p\left(\frac{X_i-c}{h}\right)\br_s\left(\frac{X_i-c}{h}\right)^\prime\right\}\right] \\
        &=\int\limits_{0}^\infty \bmu_{W}(uh+c)^\prime\otimes \br_p\left(u\right)\br_s\left(u\right)^\prime K(u)f(uh+c)\de u=:\widetilde{\bG}_{+,2}(h),
    \end{align*}
    The generic element of $\widehat{\bG}_{+,2}(h)$ is
    \[\frac{1}{nh}\sum_{i=1}^n\indic(X_i\geq c) W_{i\ell}\cdot\left(\frac{X_i-c}{h}\right)^j K\left(\frac{X_i-c}{h}\right), \qquad \ell\in\{1,\cdots,d\},\:\:j\in\{0,\cdots,p+s\}.\]
    Fix $\ell\in\{1,\cdots,d\} $ and $j\in\{0,\cdots,p+s\}$. Then
    \begin{align*}
        &\V\left[\frac{1}{nh}\sum_{i=1}^n\indic(X_i\geq c) W_{i\ell}\cdot\left(\frac{X_i-c}{h}\right)^j K\left(\frac{X_i-c}{h}\right)\right] \\
        &\qquad\qquad =\frac{1}{nh^2}\V\left[\indic(X_i\geq c) W_{i\ell}\cdot\left(\frac{X_i-c}{h}\right)^j K\left(\frac{X_i-c}{h}\right)\right]\tag{Assumption \ref{ass: sampling}}\\
        &\qquad\qquad \leq \frac{1}{nh^2}\E\left[\indic(X_i\geq c) W_{i\ell}^2\cdot\left(\frac{X_i-c}{h}\right)^{2j} K\left(\frac{X_i-c}{h}\right)^2\right]\\
        &\qquad\qquad  =\frac{1}{nh}\int\limits_0^\infty \mu_{W_\ell^2}(uh+c) u^{2j} K(u)^2 f(uh+c)\de u \tag{$u=(x-c)/h$}  = O(n^{-1}h^{-1}),
    \end{align*}
    where the last equality follows because of Assumptions \ref{ass: density running variable}, \ref{ass: kernel}, and the fact that $\mu_{W_\ell^2}(\cdot)$ is bounded in a neighborhood of $c$, which is granted by Assumption \ref{ass: SRD continuous W} and $\mathsf{k}\geq h$. Hence, by Lemma \ref{lemma: markov}
    \[\widehat{\bG}_{+,2}(h) = \widetilde{\bG}_{+,2}(h) + \Op(1/\sqrt{nh}).\]
    Consider $\widehat{\bG}_{+,3}(h)$. Using a similar logic as above 
    \begin{align*}
     \E[\widehat{\bG}_{+,3}(h)]&=\frac{1}{h}\E\left[ \bmu_{WW}(X_i)\otimes\indic ( X_i\geq c)K\left(\frac{X_i-c}{h}\right)\br_s\left(\frac{X_i-c}{h}\right)\br_s\left(\frac{X_i-c}{h}\right)^\prime\right]\\
     &=\int\limits_{0}^\infty \bmu_{WW}(uh+c)\otimes K\left(u\right)\br_s\left(u\right)\br_s\left(u\right)^\prime f(uh+c)\de u =: \widetilde{\bG}_{+,3}(h)
    \end{align*}
    and for $\ell, \ell' \in\{1,\cdots,d\} $ and $j\in\{0,\cdots,2s\}$. Then
    \begin{align*}
        &\V\left[\frac{1}{nh}\sum_{i=1}^n\indic(X_i\geq c) W_{i\ell}W_{i\ell'}\cdot\left(\frac{X_i-c}{h}\right)^j K\left(\frac{X_i-c}{h}\right)\right] \\
        &\qquad\qquad =\frac{1}{nh^2}\V\left[\indic(X_i\geq c) W_{i\ell}W_{i\ell'}\cdot\left(\frac{X_i-c}{h}\right)^j K\left(\frac{X_i-c}{h}\right)\right]\tag{Assumption \ref{ass: sampling}}\\
        &\qquad\qquad \leq \frac{1}{nh^2}\E\left[\indic(X_i\geq c) W_{i\ell}^2W_{i\ell'}^2\cdot\left(\frac{X_i-c}{h}\right)^{2j} K\left(\frac{X_i-c}{h}\right)^2\right]\\
        &\qquad\qquad  =\frac{1}{nh}\int\limits_0^\infty \E[W_{i\ell}^2W_{i\ell'}^2\mid X_i=uh+c] u^{2j} K(u)^2 f(uh+c)\de u \tag{$u=(x-c)/h$}  = O(n^{-1}h^{-1}),
    \end{align*}
    where the last equality follows because of Assumptions \ref{ass: density running variable}, \ref{ass: kernel}, and the fact that $\E[W_{i\ell}^2W_{i\ell'}^2\mid X_i=uh+c]$ is bounded in $[0,c+h]$, which is granted by Assumption \ref{ass: SRD continuous fourth moment of W} and $\mathsf{k}\geq h$. 
    
    Last, we prove that $\widehat{\bGamma}_{+,p,s}(h)$ is asymptotically invertible. To do so, note that by taking the limit as $h\to 0$, the continuity of $f(\cdot)$ (Assumption \ref{ass: density running variable}), $\bmu_W(\cdot)$, and $\bmu_{WW}(\cdot)$ (Assumption \ref{ass: SRD continuous W}) give us that
    \begin{align*}
        \widetilde{\bLambda}_{+,p,p}(h) &= f(c)\int\limits_0^\infty K(u) \br_p(u)\br_p(u)^\prime \de u + o(1) = \bLambda_{+,p,p} + o(1), \\
        \widetilde{\bG}_{+,2}(h) &= \bmu_{W}^\prime \otimes f(c)\int\limits_0^\infty K(u) \br_p(u)\br_s(u)^\prime \de u + o(1)=\bmu_{W}^\prime \otimes \bLambda_{+,p,s} + o(1) = \bG_{+,3}+o(1), \\
        \widetilde{\bG}_{+,3}(h) &= \bmu_{WW} \otimes f(c)\int\limits_0^\infty K(u) \br_s(u)\br_s(u)^\prime \de u + o(1)= \bmu_{WW}\otimes \bLambda_{+,s,s} + o(1) = \bG_{+,3} + o(1).
    \end{align*}
    Therefore, 
    \[\widehat{\bGamma}_{+,p,s}(h) = \bGamma_{+,p,s} + o(1) + \Op(1/\sqrt{nh}),\quad\text{where}\quad \bGamma_{+,p,s} = \begin{bmatrix}
        \bLambda_{+,p,p} & \bG_{+,2} \\
        \bG_{+,2}' & \bG_{+,3}
    \end{bmatrix},\]
    which is non-singular by Assumption \ref{ass: pd gram matrix}.
\end{proof}

\subsection{Proof of Lemma \ref{lemma: vartheta hat order}}
\begin{proof}\label{proof: vartheta hat order}
First, 
\begin{align*}
    \E[\widehat{\bzeta}_{+,p,s,a}(h)] &= \E\left[\frac{1}{nh}\sum_{i=1}^n\br_{p,s}\left(\frac{X_i-c}{h},\bW_i\right)K\left(\frac{X_i-c}{h}\right)\left(\frac{X_i-c}{h}\right)^{a+1}\right] \\
    &= \int\limits_0^\infty K(u)\br_{p,s}(u,\bmu_{W}(uh+c))u^{a+1} f(c + hu)\de u \\
    &= f(c)\int\limits_0^\infty K(u)\br_{p,s}(u,\bmu_{W})u^{a+1} \de u + O(h).
\end{align*}
Moreover, the generic element of $\widehat{\bzeta}_{+,p,s,a}(h)$ is either
\begin{align*}
\frac{1}{nh}\sum_{i=1}^n\indic(X_i\geq c)K\left(\frac{X_i-c}{h}\right)\left(\frac{X_i-c}{h}\right)^{j+a+1}, \qquad j\in\{0,1,\cdots, p\},
\end{align*}
or
\begin{align*}
\frac{1}{nh}\sum_{i=1}^n\indic(X_i\geq c)K\left(\frac{X_i-c}{h}\right)W_{i\ell}\cdot\left(\frac{X_i-c}{h}\right)^{j+a+1}, \qquad j\in\{0,1,\cdots, s\},\:\:\ell\in\{1,\cdots, d\}.
\end{align*}
The variances of these elements are of order $O(n^{-1}h^{-1})$. Indeed, the variance of the latter is
\begin{align*}
  \V&\left[\frac{1}{nh}\sum_{i=1}^n\indic(X_i\geq c)K\left(\frac{X_i-c}{h}\right)W_{i\ell}\cdot\left(\frac{X_i-c}{h}\right)^{j+a+1}\right]\\ 
  &\qquad\leq \frac{1}{nh^2}\E\left[\indic(X_i\geq c)K^2\left(\frac{X_i-c}{h}\right)W_{i\ell}^2\cdot\left(\frac{X_i-c}{h}\right)^{2(j+a+1)}\right] \\
  &\qquad=\frac{1}{nh}\int\limits_0^\infty \mu_{W_\ell}(uh+c)u^{2(j+a+1)}K^2(u)f(uh+c) = O\left(\frac{1}{nh}\right),
\end{align*}
where for the last equality we used Assumptions \ref{ass: sampling} and \ref{ass: kernel} and $\mathsf{k}\geq h$. Then, $\widehat{\bzeta}_{+,p,s,a}(h) = \E[\widehat{\bzeta}_{+,p,s,a}(h)] + \Op(1/\sqrt{nh})$ by Lemma \ref{lemma: markov}. A similar argument goes through for $\widehat{\bvarphi}_{+,p,s,a}(h)$.

The last part of the lemma follows by taking the limit as $h\to 0$ and using the continuity of $f(\cdot)$ (Assumption \ref{ass: density running variable}), $\bmu_W(\cdot)$, and $\bmu_{WW}(\cdot)$ (Assumption \ref{ass: SRD continuous W}).
\end{proof}

\subsection{Proof of Lemma \ref{lemma: first order variance}}
\begin{proof}\label{proof: first order variance}
Fix $h>0$ and define
\begin{align*}
    \bw_{+,i}(h):=\indic(X_i\geq c)K\left(\frac{X_i-c}{h}\right)\br_{p,s}\left(\frac{X_i-c}{h},\bW_i\right),
\end{align*}
so that we can write
\[\bL_{+,p,s}(h)= \frac{1}{nh}\sum_{i=1}^n\bw_{+,i}(h)u_{+,i}(h).\]
By the properties of $\ell_2$-residuals, we get that $\E[\bw_{+,i}(h)u_{+,i}(h)]=0$ and so $\E[\bL_{+,p,s}(h)]=0$. The variance of $\bL_{+,p,s}(h)$ is 
\begin{align*}
     \V\left[\frac{1}{nh}\sum_{i=1}^n\bw_{+,i}(h)u_{+,i}(h)\right] &= \frac{1}{n^2h^2}\sum_{i=1}^n\sum_{j=1}^n\Cov(\bw_{+,i}(h)u_{+,i}(h), \bw_{+,j}(h)u_{+,j}(h)) \\
    &= \frac{1}{nh^2}\V[\bw_{+,i}(h)u_{+,i}(h)] \tag{Assumption \ref{ass: sampling}} \\
    &= \frac{1}{nh}\underbrace{\frac{1}{h}\E[\bw_{+,i}(h)\bw_{+,i}(h)^\prime u^2_{+,i}(h)]}_{:=\bV_{+,p,s}(h)} \tag{$\E[\bw_{+,i}(h)u_{+,i}(h)]=0$}.
\end{align*} 
    
By the law of iterated expectations and changing variables, we get\footnotesize
\begin{align*}
    &\mathbf{V}_{+,p,s}(h) =\\
    &\quad \int\limits_0^\infty f(uh+c) K^2(u)\begin{bmatrix}
        \br_p\left(u\right)\br_p\left(u\right)^\prime \E[u_{+,i}^2(h)\mid X_i=uh+c] & \E[\bW_i^\prime u_{+,i}^2(h)\mid X_i=uh+c]  \otimes \br_p\left(u\right)\br_s\left(u\right)^\prime \\
        \E[\bW_i u_{+,i}^2(h)\mid X_i=uh+c]  \otimes \br_s\left(u\right)\br_p\left(u\right)^\prime & \E[\bW_i\bW_i^\prime u_{+,i}^2(h)\mid X_i=uh+c]  \otimes \br_s\left(u\right)\br_s\left(u\right)^\prime
    \end{bmatrix} \de u,
\end{align*}\normalsize

Finally, the last part of the lemma follows by taking the limit as $h\to 0$, using the properties of compact kernels (Assumption \ref{ass: kernel}), the continuity of $f(\cdot)$ (Assumption \ref{ass: density running variable}), and the bounded conditional fourth moments of $(Y_i,\bW_i')'$ (Assumptions \ref{ass: continuous fourth moment of Y}-\ref{ass: SRD continuous fourth moment of W}),   together with the Cauchy-Schwarz inequality and $\mathsf{k}\geq h$. 
\end{proof}

\subsection{Proof of Theorem \ref{thm: asymptotic distribution}}
\begin{proof}\label{proof: asymptotic distribution}
    Fix an arbitrary $\bc\in\R^{1+p+d}\setminus \{\mathbf{0}_{1+p+d}\}$. Consider 
    $$\phi_{+,p,s}(h) := \sqrt{nh}\bc^\prime \bH_{p,s}(h)(\widehat{\bvartheta}_{+,p,s}(h) - \bvartheta_{+,p,s}^\star(h))=  \sqrt{nh}\bc^\prime\widehat{\bGamma}_{+,p,s}^{-1}(h)\bL_{+,p,s}(h).$$
    By Lemma \ref{lemma: grand hessian} and Lemma \ref{lemma: first order variance}, we get
    \[\phi_{+,p,s}(h) = \dot{\phi}_{+,p,s}(h) +\op(1), \qquad \dot{\phi}_{+,p,s}(h)= \sqrt{nh}\bc^\prime\widetilde{\bGamma}_{+,p,s}^{-1}(h)\bL_{+,p,s}(h).\]
    Then, by Lemma \ref{lemma: first order variance}
    \begin{align*}
        \V[\sqrt{nh}\bc^\prime\widetilde{\bGamma}_{+,p,s}^{-1}(h)\bL_{+,p,s}(h)] &= \bc^\prime\bOmega_{+,p,s}\bc+ o(1), \qquad\bOmega_{+,p,s}:={\bGamma}_{+,p,s}^{-1}\mathbf{V}_{+,p,s}
    {{\bGamma}_{+,p,s}^{-1\prime}}.
    \end{align*}    
    Positive definiteness of $\bOmega_{+,p,s}$ follows from the fact that it is a product of symmetric positive definite matrices (Lemma \ref{lemma: grand hessian} and Lemma \ref{lemma: first order variance}). Finally, by appropriately scaling $\dot{\phi}_{+,p,s}(h)$ by its variance, we can rewrite it as
    \[\sum_{i=1}^nw_{+,i}(h)u_{+,i}(h),\]
    with
    \[w_{+,i}(h) := \left(\bc^\prime\bOmega_{+,p,s}\bc\right)^{-1/2}\bc^\prime\widetilde{\bGamma}_{+,p,s}^{-1}(h)\indic(X_i\geq c)K\left(\frac{X_i-c}{h}\right)\br_{p,s}\left(\frac{X_i-c}{h},\bW_i\right)/\sqrt{nh}.\]
        
    Therefore, $\{{w}_{+,i}(h)u_{+,i}(h)\}_{i=1}^n$ is a triangular array of mean-zero row-wise independent summands with variance converging to 1 to which we can apply a Lindeberg-Feller CLT. To do so, we verify the Lindeberg condition by showing that the stronger Lyapunov condition holds for the fourth moment. To see this, by Assumption \ref{ass: DGP}
    \begin{align*}
        \sum_{i=1}^n\E[|{w}_{+,i}(h)u_i(h)|^4]&\lesssim \frac{1}{nh^2}\cdot \int\limits_c^\infty \left|\cbe_{\nu}^\prime\widetilde{\bGamma}_{+,p,s}^{-1}(h)K\left(\frac{x-c}{h}\right)\br_{p,s}\left(\frac{x-c}{h},\bmu_{W}(x)\right)\right|^4f(x)\de x = O\left(\frac{1}{nh}\right),
    \end{align*}    
    where the final order comes from the change of variables $u=\frac{x-c}{h}$ and the fact that the resulting integral is bounded by Assumptions \ref{ass: density running variable}, \ref{ass: DGP}, and \ref{ass: kernel}. Therefore, we can conclude that
    \[\sqrt{nh}\bc^\prime\bH_{p,s}(h)\widehat{\bGamma}_{+,p,s}^{-1}(h)\bL_{+,p,s}(h) \cdist \mathsf{N}(0, \bc^\prime\bOmega_{+,p,s}\bc).\]
    As $\bc$ was chosen arbitrarily, by the Cram\'{e}r-Wold device we get 
    \[\sqrt{nh}\bH_{p,s}(h)\widehat{\bGamma}_{+,p,s}^{-1}(h)\bL_{+,p,s}(h) \cdist \mathsf{N}(\mathbf{0}_{1+p+d}, \bOmega_{+,p,s}),\]
    which was to be shown. 
\end{proof}

\subsection{Proof of Theorem \ref{theorem: variance estimation plug-in}}
\begin{proof}\label{proof: variance estimation plug-in}
    First, note that under the assumptions of Theorem \ref{thm: asymptotic distribution}, we get that 
    \begin{align}\label{eq: vanishing leverage}
        \max_{i\in[n]}L_{-,i}=\op(1)\qquad\text{and}\qquad \max_{i\in[n]}L_{+,i}=\op(1).
    \end{align}
    The result above follows from Assumptions \ref{ass: sampling}, \ref{ass: DGP}, \ref{ass: kernel}, and Lemma \ref{lemma: grand hessian}.

    Define $\omega_{+,i}(h):= \indic(X_i\geq 0)K\left(\frac{X_i-c}{h}\right)\sqrt{w_{+,i}(h)}$. First, note that
    \[\widehat{u}_{+,i}(h) = u_{+,i}(h) + \widehat{v}_{+,i}(h),\]
    where $u_{+,i}(h):= Y_i - \br_{p,s}(X_i-c,\bW_i)^\prime\bvartheta^\star_{+,p,s}(h)$ and $\widehat{v}_{+,i}(h):= \br_{p,s}(X_i-c,\bW_i)^\prime({\bvartheta}^\star_{+,p,s}(h)- \widehat{\bvartheta}_{+,p,s}(h))$. Then, focusing on the ``meat'' part of the estimator, we get
    \begin{align*}
        \widehat{\mathbf{V}}_{+,p,s}(h) &= \frac{1}{nh}\sum_{i=1}^n\omega_{+,i}(h)\br_{p,s}\left(\frac{X_i-c}{h},\bW_i\right)\br_{p,s}\left(\frac{X_i-c}{h},\bW_i\right)^\prime {\widehat{u}_{+,i}}^2(h)  \\
        &= \frac{1}{nh}\sum_{i=1}^n\omega_{+,i}(h)\br_{p,s}\left(\frac{X_i-c}{h},\bW_i\right)\br_{p,s}\left(\frac{X_i-c}{h},\bW_i\right)^\prime u_{+,i}^2(h) \tag{$:=\bV_{+,1}(h)$}  \\
        &\quad + \frac{1}{nh}\sum_{i=1}^n\omega_{+,i}(h)\br_{p,s}\left(\frac{X_i-c}{h},\bW_i\right)\br_{p,s}\left(\frac{X_i-c}{h},\bW_i\right)^\prime \widehat{v}_{+,i}^2(h) \tag{$:=\bV_{+,2}(h)$}, \\
        &\quad + \frac{2}{nh}\sum_{i=1}^n\omega_{+,i}(h)\br_{p,s}\left(\frac{X_i-c}{h},\bW_i\right)\br_{p,s}\left(\frac{X_i-c}{h},\bW_i\right)^\prime \widehat{v}_{+,i}(h){u}_{+,i}(h) \tag{$:=\bV_{+,3}(h)$}.
    \end{align*}
    
     For convenience, let $\bV_{+,1}^{\mathtt{HC0}}(h)$ denote $\bV_{+,1}(h)$ when $w_{+,i}(h) = 1$. Note that $\bV^{\mathtt{HC0}}_{+,1}(h) \cprob \mathbf{V}_{+,p,s}$ by the weak law of large numbers. Then, it is immediate to see that $\bV_{+,1}^{\mathtt{HC1}}(h) = \frac{n-k}{n}\bV_{+,1}^{\mathtt{HC0}}(h)$ and so $\bV_{+,1}^{\mathtt{HC1}}(h)\cprob \mathbf{V}_{+,p,s}$. Regarding the HC2-type estimator, it suffices to note that
     \[|\bV_{+,1}^{\mathtt{HC2}}(h)-\bV_{+,1}^{\mathtt{HC0}}(h)| \leq \left(\frac{1}{nh}\sum_{i=1}^n\left|\br_{p,s}\left(\frac{X_i-c}{h},\bW_i\right)\right|^2 u_{+,i}^2(h)\right) \cdot \left|\frac{1}{1-\max_{i\in[n]}L_i} - 1\right|,\]
     where the first term is $\Op(1)$ by Assumptions \ref{ass: sampling}-\ref{ass: DGP} and the second term is $\op(1)$ due to \eqref{eq: vanishing leverage}. A similar argument works for the HC3-type estimator. 
     
     At this point, if we show that $\bV_{+,j}(h)\cprob 0, j=2,3,$ then we have established the consistency of our estimator for the variance of the RD estimator. Let's start with $\bV_{+,2}(h)$:
    \begin{align*}
        \left|\bV_{+,2}(h)\right| &\leq 
         {\frac{1}{nh}\sum_{i=1}^n|\omega_{+,i}(h)| \left|\br_{p,s}\left(\frac{X_i-c}{h},\bW_i\right)\right|^2\left|\br_{p,s}\left(\frac{X_i-c}{h},\bW_i\right)\right|^2\left|\bH_{p,s}(h)\left(\widehat{\bvartheta}_{+,p,s}(h)-\bvartheta^\star_{+,p,s}(h)\right)\right|^2} \\
         &\leq \left|\bH_{p,s}(h)\left(\widehat{\bvartheta}_{+,p,s}(h)-\bvartheta^\star_{+,p,s}(h)\right)\right|^2\cdot\max_{i\in[n]}\left|\omega_{+,i}(h)\right|\cdot\frac{1}{nh}\sum_{i=1}^n\left|\br_{p,s}\left(\frac{X_i-c}{h},\bW_i\right)\right|^4.
    \end{align*}
    The first term on the right-hand side is $\Op(n^{-1}h^{-1})$ by Theorem \ref{thm: asymptotic distribution}, the second term is $\Op(1)$ by \eqref{eq: vanishing leverage}, whereas the third term is $\Op(1)$ as it converges in probability to its expectation which is
    \[\int\limits_0^\infty f(uh+c)K(u) \left|\br_{p,s}(u,\bmu_{W}(uh+c))\right|^4\de u,\]
    which is finite because of Assumptions \ref{ass: density running variable}, \ref{ass: DGP}, and \ref{ass: kernel}. Therefore, $\bV_{+,2}(h) = \op(1)$.
    
    Then
    \begin{align*}
        |\bV_{+,3}(h)|
        &\leq \left|\bH_{p,s}(h)\left(\widehat{\bvartheta}_{+,p,s}(h)-\bvartheta_{+,p,s}^\star(h)\right)\right| \cdot\max_{i\in[n]}\left|\omega_{+,i}(h)\right|\cdot \frac{1}{nh}\sum_{i=1}^n\left|\br_{p,s}\left(\frac{X_i-c}{h},\bW_i\right)\right|^3 |u_{+,i}(h)|.
    \end{align*}
    The first term on the right-hand side is $\Op(1/\sqrt{nh})$ by Theorem \ref{thm: asymptotic distribution}. The second term is $\Op(1)$ by\eqref{eq: vanishing leverage}. To see that the third term is $\Op(1)$, note that by the H{\"o}lder's inequality 
    \[\E\left[\left|\br_{p,s}\left(\frac{X_i-c}{h},\bW_i\right)\right|^3 |u_{+,i}(h)|\right] \leq \left(\int\limits_0^\infty f(uh+c)K(u)\left|\br_{p,s}\left(u,\bmu_{W}(uh+c)\right)\right|^3 \de u\right)^{3/4} \E[u_{+,i}^4(h)]^{1/4},\]
    which is finite because of Assumptions \ref{ass: density running variable}, \ref{ass: DGP}, and \ref{ass: kernel}. Therefore, $\bV_{+,3}(h) = \Op(1/\sqrt{nh})=\op(1)$. 

    In conclusion, we showed that
    \[\widehat{\mathbf{V}}_{+,p,s}(h) = \sum_{\ell=1}^3\bV_{+,\ell}(h),\]
    where $\bV_{+,1}(h)$ converges in probability to $\mathbf{V}_{+,p,s}(h)$ and the other terms are all $\op(1)$. Moreover, Lemma \ref{lemma: grand hessian} gives us $\widehat{\bGamma}_{+,p,s}^{-1}(h)\cprob {\bGamma}_{+,p,s}^{-1}.$ Hence, by Slutsky's theorem
    \[\widehat{\mathcal{V}}_{+,\nu,p,s}(h) \cprob {\mathcal{V}}_{+,\nu,p,s}. \qedhere \]
\end{proof}

\subsection{Proof of Lemma \ref{lemma: smoothing bias}}
\begin{proof}\label{proof: smoothing bias}
    Let $q:=p\vee s$. First of all, if we take a $(q +2)$-th order Taylor expansion of $\mu_+(\cdot,\bw)$ around $c$, we get
    \begin{align*}
      {\bvartheta}^\star_{+,p,s}(h)&= \bH_{p,s}^{-1}(h)\widetilde{\bGamma}_{+,p,s}^{-1}(h)\E\left[\frac{1}{h}\indic(X_i\geq c)K\left(\frac{X_i-c}{h}\right)\br_{p,s}\left(\frac{X_i-c}{h},\bW_i\right)\mu(X_i,\bW_i)\right]\\
      &=\bvartheta_{+,p,s} +\bH_{p,s}^{-1}(h)\widetilde{\bGamma}_{+,p,s}(h)^{-1} \left(h^{p+1}\widetilde{\bzeta}_{+,p,s,p}(h) \frac{\alpha^{(p+1)}_{+}(c)}{(p+1)!} + h^{p+2}\widetilde{\bzeta}_{+,p,s,p+1}(h) \frac{\alpha^{(p+2)}_{+}(c)}{(p+2)!} \right) \\
      &\:\:+ \bH_{p,s}^{-1}(h)\widetilde{\bGamma}_{+,p,s}(h)^{-1} \left(h^{s+1}\widetilde{\bvarphi}_{+,p,s,s}(h) \frac{\blambda^{(s+1)}_{+}(c)}{(s+1)!} + h^{s+2}\widetilde{\bvarphi}_{+,p,s,s+1}(h) \frac{\blambda^{(s+2)}_{+}(c)}{(s+2)!} \right) \\
      &\:\:+ o(h^{2+p\wedge s}),
    \end{align*}
    where 
    \begin{align*}
    \widetilde{\bzeta}_{+,p,s,a}(h)&=\int\limits_0^\infty K(u)\br_{p,s}(u,\bmu_{W}(uh+c))u^{a+1} f(uh+c)\de u,\\
    \widetilde{\bvarphi}_{+,p,s,a}(h)&=\int\limits_0^\infty K(u) \begin{bmatrix}
    \br_p(u)\bmu_{W}(uh+c)^\prime \\
    \bmu_{WW}(uh+c)\otimes \br_s(u)
\end{bmatrix} u^{a+1} f(uh+c)\de u.       
    \end{align*}
    The $o(h^{2+p\wedge s})$ term comes from the integral form of the remainder of the Taylor expansion. Suppose $p=s$. The remainder is of the form
    \begin{align*}
        \bv_{+}(h)=\int\limits_0^\infty\int\limits_{\R^d} K\left(u\right)\br_{p,s}\left(u,\bw\right)\int\limits_c^{uh+c}\Bigg(\frac{\mu_+^{(p+3)}(t, \bw)}{(p+2)!}(uh+c-t)^{p+2}\Bigg)\de tf_{X,W}(uh+c,\bw)\de \bw\de u.
    \end{align*}
    Being $\mu_+^{(p+3)}(x,\bw)$ continuous by Assumption \ref{ass: SRD continuous Y}, $\mathsf{k}\geq h$, and using Assumptions \ref{ass: density running variable} and \ref{ass: kernel}, we get $\bv_+(h) = o(h^{2+p})$. 
    
    Whenever $s\neq p$, the remainder $\bv_+(h)$ includes additional terms we ignored in the linearization above. These terms will always include either $\alpha^{(\ell)}(c), \ell \geq p + 3$ or $\lambda^{(\ell)}(c), \ell \geq s + 3$ and be of order $o(h^{2+p})$ and $o(h^{2+s})$, respectively.

    The last part of the lemma follows by taking the limit as $nh\to\infty$ and $h\to 0$ and relying on Lemma \ref{lemma: grand hessian} and \ref{lemma: vartheta hat order}.
\end{proof}

\bibliography{CCFPT_2025_HTERD--bib}
\bibliographystyle{jasa}
